\documentclass[usenatbib, twocolumn, nofootinbib]{aastex631}

\usepackage{multirow}
\usepackage{bigints}
\usepackage{natbib}
\usepackage{color}
\usepackage{wrapfig}
\usepackage{amsmath,amsfonts,bm}
\usepackage{hyperref}
\usepackage[T1]{fontenc}
\usepackage{graphicx}	
\usepackage[mathlines]{lineno}
\usepackage{wrapfig}
\usepackage{fontawesome}

\newcolumntype{C}[1]{>{\centering\let\newline\\\arraybackslash\hspace{0pt}}m{#1}}

\newcommand{\srini}[1]{\textcolor{red}{[SR: #1]}}

\newcommand{\pending}[1]{\textcolor{blue}{(#1)}}

\newcommand{\cmbsfour}{CMB-S4}
\newcommand{\sfour}{S4-Wide}
\newcommand{\sfourdeep}{S4-Ultra Deep}

\newcommand{\spt}{SPT}

\newcommand{\sptthreeg}{SPT-3G}
\newcommand{\sptfour}{SPT-3G+}

\newcommand{\planck}{{\it Planck}}
\newcommand{\litebird}{LiteBIRD}

\newcommand{\snr}{S/N}

\newcommand{\sogoal}{SO-Goal}
\newcommand{\sofid}{SO-Baseline}
\newcommand{\ccatprime}{FYST}
\newcommand{\agora}{\textsc{Agora}}
\newcommand{\websky}{Websky}
\newcommand{\amber}{AMBER}

\newcommand{\taure}{\tau_{\rm re}}
\newcommand{\omchsq}{\Omega_{c}h^{2}}
\newcommand{\ombhsq}{\Omega_{b}h^{2}}

\newcommand{\fskywideclean}{50\%}

\newcommand{\fskywidefull}{67\%}
\newcommand{\fskydeepfull}{3\%}
\newcommand{\fskyso}{40\%}

\newcommand{\ukam}{\ensuremath{\mu}{\rm K{\text -}arcmin}}
\newcommand{\uk}{\ensuremath{\mu}{\rm K}}
\newcommand{\uksq}{\ensuremath{\mu}{\rm K}^{2}}
\newcommand{\sqdeg}{{\rm deg}^{2}}
\newcommand{\mjy}{{\rm mJy}}

\newcommand{\msol}{\ensuremath{\mbox{M}_{\odot}}}

\newcommand{\lcdm}{\Lambda{\rm CDM}}

\newcommand{\comment}[1]{}
\newcommand{\sz}{Sunyaev-Zel{'}dovich}

\newcommand{\clustermask}{2 \times 10^{14}}

\newcommand{\thetavir}{\theta_{500c}}
\newcommand{\mvir}{M_{\rm 500c}}

\newcommand{\dl}{D_{\ell}}

\newcommand{\kszleveldl}{3}
\newcommand{\ellmax}{\ell_{\rm max}}
\newcommand{\aksz}{A_{\rm kSZ}}

\newcommand{\alpharad}{\alpha_{\rm rad}}
\newcommand{\alpharadsigma}{\sigma(\alpha_{\rm rad})}
\newcommand{\ellmaxksz}{\ell_{\rm max}^{\rm TT, kSZ}}
\newcommand{\acibtsz}{A_{\rm CIB+tSZ}}
\newcommand{\zmid}{z_{\rm re}^{\rm mid}}
\newcommand{\zdur}{\Delta z_{\rm re}}
\newcommand{\akszhomo}{A_{\rm kSZ}^{h}}
\newcommand{\alphakszhomo}{\alpha_{\rm kSZ}^{h}}
\newcommand{\tszfree}{tSZ-free}
\newcommand{\cibfree}{CIB-free}

\newcommand{\clcov}{{\bf C}_{\rm \ell}}
\newcommand{\clinv}{\clcov^{-1}}

\newcommand{\abstracttext}{
We propose a cross-internal linear combination (cross-ILC) approach to measure the small-scale cosmic microwave background (CMB) anisotropies robustly against the contamination from astrophysical signals. In particular, we focus on the mitigation of systematics from cosmic infrared background (CIB) and thermal \sz{} (tSZ) signals in kinematic \sz{} (kSZ) power spectrum and CMB lensing.
We show the cross-spectrum measurement between two CMB maps created by nulling the contributions from CIB (\cibfree{} map) and tSZ (\tszfree{} map) to be robust for kSZ as the approach significantly suppresses the total contribution of CIB and tSZ signals. 
Similarly, for CMB lensing, we use the approach introduced by \citet{madhavacheril18} but with a slight modification by using the \tszfree{} and \cibfree{} maps in the two legs of the quadratic estimator. 
By cross-correlating the CMB lensing map created using this technique with galaxy surveys, we show that the biases from both CIB/tSZ are negligible. 
We also compute the impact of unmodeled CIB/tSZ residuals on kSZ and cosmological parameters finding that the kSZ measured using the standard ILC to be significantly biased. 
The kSZ estimate from the cross-ILC remains less affected by CIB/tSZ making it crucial for current and upcoming CMB surveys such as the South Pole Telescope (SPT), Simons Observatory (SO) and CMB-S4.
With the cross-ILC approach, we find the total kSZ power spectrum can be measured at very high significance in the coming years: $35\sigma$ by SPT, $22\sigma$ by SO, and $80\sigma$ by CMB-S4. 
Finally, we forecast constraints on the epoch of reionization using the kSZ power spectrum and find that the duration of reionization, currently unconstrained by \planck, can be constrained to $\sigma(\zdur)$ = 1.5 (or) 0.5 depending on the choice of the prior on the optical depth to reionization. 
The data products and the associated codes can be downloaded from this \href{https://github.com/sriniraghunathan/cross_ilc_methods_paper}{link$^{\text{\faGithub}}$}.
}

\newcommand{\tittext}{A Cross-Internal Linear Combination Approach to Probe the Secondary CMB Anisotropies: Kinematic \sz{} Effect and CMB Lensing}

\defcitealias{madhavacheril18}{MH18}
\defcitealias{reichardt21}{R21}

\begin{document}

\title{\tittext}
\newcommand{\shortauthourlist}{Raghunathan \& Omori}
\shorttitle{Cross-ILC technique for secondary CMB anisotropies}
\shortauthors{\shortauthourlist}

\author[0000-0003-1405-378X]{Srinivasan Raghunathan} 
\affiliation{Center for AstroPhysical Surveys, National Center for Supercomputing Applications, Urbana, IL 61801, USA}

\author[0000-0002-0963-7310]{Yuuki Omori}
\affiliation{Department of Astronomy and Astrophysics, University of Chicago, Chicago, IL 60637, USA}
\affiliation{Kavli Institute for Cosmological Physics, University of Chicago, Chicago, IL 60637, USA}

\begin{abstract}
\abstracttext{}
\end{abstract}

\correspondingauthor{Srinivasan Raghunathan}\email{srinirag@illinois.edu}
\correspondingauthor{Yuuki Omori}\email{yomori@uchicago.edu}

\section{Introduction}
\label{sec_introduction}


Observations of the millimeter sky offer a wealth of information about the origin, contents, and evolution of our Universe as well as astrophysical effects. In this work, we focus on two late-time effects: gravitational lensing and the Sunyaev Zel'dovich (SZ) effect. The effect of gravitational lensing -- the bending of the path of the CMB photons due to the presence of matter in the Universe -- is an excellent probe of structure growth.
The SZ can be of two kinds: kinematic (kSZ, \citealt{sunyaev80b}) and thermal (tSZ, \citealt{sunyaev70, sunyaev72}) effects. 
The tSZ arises from the inverse Compton scattering of CMB photons off free electron in hot intracluster medium (ICM). 
As a result, it is a strong probe of both structure formation and the complex gas physics of the ICM.
The kSZ signal originates when moving electrons with bulk motion scatter the CMB photons introducing a Doppler shift. 
This can arise due to spatial differences in ionization fractions in the high redshift ($z \ge 6$) universe during the epoch of reionization (EoR, \citealt{paul21}) and also due to bulk motion of haloes containing free electrons in the post-reionization Universe. 
From now on, we will refer to these two as patchy-kSZ and homogeneous-kSZ signals. 
These two kSZ signals can provide key information on the physics of the EoR \citep[see][for a review]{choudhury22} and the growth of structure, respectively.

Both CMB lensing \citep{smith07, das11, vanengelen12, planck14_lensing, story14, omori17, wu19, madhavacheril23} and tSZ \citep{plancksz15, madhavacheril20, bleem21} have been detected to very high significance by many experiments. 
They have also been used to constrain cosmology and astrophysics \citep{plancksz15, bianchini20, douspis21, sanchez22, omori23, qu23}. 
In addition, the tSZ signal has also been used to produce catalogs of tens of thousands of galaxy clusters \citep{bleem15, planckSZcat16, hilton20, huang20} which have also been used as cosmological probes \citep[][for example]{plancksz15, bocquet19}.
Although the kSZ signal has also been detected \citep{hand12, soergel16, hill16, reichardt21, gorce22}, the significance of these detections has not been high enough for them to be used as cosmological or EoR probes \citep{chen23}. 
Of these, \citet{reichardt21} and \citet{gorce22} measured the total kSZ power spectrum which receives contribution from both the patchy and homogeneous terms while the other works focussed only on the homogeneous term due to the motion of haloes. 
The improvement in the quality of CMB data from current and future surveys \citep{benson14, bender18, SO18, cmbs4collab19, sehgal19}  significantly improves the expected detection significance of the lensing and SZ signals. 
However, with the improvement in the statistical errors, the systematic errors in the measurements also become important and non-negligible.

In this work, we focus on systematic biases in the kSZ power spectrum and CMB lensing reconstruction due to the presence of unwanted astrophysical  foreground signals in the CMB temperature maps. 
For kSZ, these contaminants include tSZ and signals from dusty star-forming galaxies (DSFGs, the integrated light from which makes up the cosmic infrared background or CIB) and radio galaxies.
For CMB lensing, all the above signals including the kSZ act as sources of bias. 

In the past, the standard approach to measure the kSZ power spectrum has been to use templates, that were predicted from simulations \citep{shaw10, shaw12, battaglia13b, battaglia13c}, for jointly fitting and mitigating the undesired  signals \citep{choi20, reichardt21, gorce22}. 
The use of templates can be harmful as a mis-estimation of the template can bias the kSZ results significantly as demonstrated in Appendix~\ref{appendix_bias_cibtsz}.
\citet{gorce22} modified the use of kSZ/tSZ templates and built an emulator to jointly predict the kSZ/tSZ signals using a random forest algorithm trained numerically for a range of parameters governing cosmology, cluster astrophysics, and reionization.
This helped them to break the degeneracy between kSZ and tSZ \citep[See Fig. 3 of][]{gorce22} which led to an almost $\times2$ improvement in the kSZ signal-to-noise (\snr). 
However, \citet{gorce22} used a template for the CIB signal. 

In the context of CMB lensing, several methods have been proposed to address the systematic errors due to foreground signals at the expense of the \snr, for example, by adopting an aggressive masking strategy \citep{vanengelen14}, by using shear-only reconstruction \citep{schaan19}, and by implementing a foreground bias-hardening technique \citep{namikawa13, osborne14, darwish21b, macCrann23}.
\citet[hereafter \citetalias{madhavacheril18}]{madhavacheril18} introduced a modified quadratic estimator (QE) \citep{hu02} to mitigate tSZ-induced biases by using a \tszfree{} map in one leg of the QE. While this lensing estimator works well to mitigate lensing bias from tSZ, the \tszfree{} maps generally have an enhanced level of CIB which might be problematic for future surveys, particularly for cross-correlation studies \citep{schmittfull18, baxter23, chang23, omori23} that rely on CMB temperature-based lensing reconstruction.

In this study, we use a different approach and propose to use the cross-power spectra between two different CMB maps constructed using different internal linear combination (ILC, \citealt{cardoso08, planck14_smica}) of the different frequency bands. 
The two CMB maps are different because they have been derived in a way to null the response of different foreground signals in them using the constrained-ILC (cILC) approach \citep{remazeilles11}. 
We refer to this cross-power spectrum approach as the {\it ``cross-ILC technique''}. 
Given that CIB and tSZ form the most important sources of biases for kSZ, the cross-ILC technique uses the \tszfree{} and \cibfree{}\footnote{CIB signals cannot be fully removed from the map using cILC as they are made up of multiple populations. As a result, a \cibfree{} map is generally a CIB-minimized map and will never be truly devoid of CIB.} CMB maps for the power spectrum estimation. 
Along these lines, we note here that \citet{kusiak23} recently explored nulling multiple foreground components using large-scale structure tracers from galaxy surveys. 
Likewise, for CMB lensing, we use the \citetalias{madhavacheril18} estimator but with a slight alteration of using \tszfree{} and \cibfree{} CMB maps in the two legs of the QE.

The paper is structured as follows: 
In \S\ref{sec_method}, we describe the experimental setups used for forecasting followed by the cross-ILC method and the simulations used for estimating the systematic errors. 
We present and discuss the results in \S\ref{sec_results} for both kSZ and CMB lensing which include the importance of systematic errors, mitigation strategy using the cross-ILC technique, expected kSZ \snr{} and constraints on the EoR. 
Finally, we discuss the potential applications of the cross-ILC technique and conclude in \S\ref{sec_conclusion}. 
In appendixes, we present the bandpower errors for kSZ; compute biases in kSZ and cosmological parameters due to residual foregrounds; discuss alternate foreground models; and demonstrate biases in the cross-correlation of CMB lensing and galaxy surveys for multiple redshift bins. 

The underlying cosmology used in this work was set to \planck{} 2018 measurements (TT, TE, EE + lowE + lensing in Table 2 of \citealt{planck20_2018cosmo}). 
For forecasting the \snr, we assume a kSZ power spectrum level of $\dl = \kszleveldl \uk^{2}$ where $\dl = \ell(\ell+1) C_{\ell} / 2\pi$.

\section{Method}
\label{sec_method}

\subsection{Experimental Specifications}
\label{sec_exp_specs}

\begin{deluxetable*}{| c || c | c | c | c | c | c | c |}
\tabletypesize{\small}
\tablecaption{Band-dependent beam and white noise levels for different experiments considered in this work. Note that we have made a conservative choice and ignored frequency channels below 90 GHz in this work.}
\label{tab_exp_specs}
\tablehead{
\multirow{2}{*}{\hspace{0.5cm}Experiment\hspace{0.5cm}} & \multicolumn{7}{c|}{Beam $\theta_{\rm FWHM}$ in arcminutes ($\Delta_{T}$ in $\ukam$)} \\
\cline{2-8}
& 90 GHz & 150 GHz & 220 GHz & 285 GHz & 345 GHz & 410 GHz & 850 GHz
}
\startdata
\hline
\sptthreeg{} & \multirow{2}{*}{1.7 (3)} & \multirow{2}{*}{1.2 (2.2)} & 1 (8.8) & \multicolumn{2}{c|}{-} & \multirow{2}{*}{-} & \multirow{2}{*}{-} \\
\cline{1-1}\cline{4-6}
+ \sptfour{} & & & 1 (2.3\tablenotemark{a}) & 0.55 (5.6) & 0.45 (40.2) & & \\\hline\hline
\sofid{} & \multirow{2}{*}{2.2 (8)} & \multirow{2}{*}{1.4 (10)} & 1 (22) & 0.9 (54) & \multicolumn{3}{c|}{-} \\\cline{1-1}\cline{4-8}
+ \ccatprime{} & & & 1 (12.3\tablenotemark{b}) & 0.8 (24.8\tablenotemark{c}) & 0.6 (107) & 0.5 (407) & 0.3 ($6.8 \times 10^{5}$) \\\hline\hline
\sogoal{} & \multirow{2}{*}{2.2 (5.8)} & \multirow{2}{*}{1.4 (6.3)} & 1 (15) & 0.9 (37) & \multicolumn{3}{c|}{-} \\\cline{1-1}\cline{4-8}
+ \ccatprime{} & & & 1 (10.6\tablenotemark{b}) & 0.8 (22.3\tablenotemark{c}) & 0.6 (107) & 0.5 (407) & 0.3 ($6.8 \times 10^{5}$) \\\hline\hline
\sfour{} & 2.2 (1.9) & 1.4 (2.09) & 1 (6.9) & 0.9 (16.88) & \multicolumn{3}{c|}{-} \\\hline
\sfourdeep{}\tablenotemark{$\dagger$} & 3.0 (0.45) & 1.92 (0.41) & 1.32 (1.3) & 1.2 (3.1) &  \multicolumn{3}{c|}{-}\\\hline\hline
\enddata
\begin{flushleft}
\tablenotetext{a}{Inverse variance weighted noise estimate of \sptthreeg{} 220 GHz and \sptfour{} 225 GHz channels.}
\tablenotetext{b}{Inverse variance weighted noise estimate the respective experiment and \ccatprime{} 285 GHz channels.}
\tablenotetext{c}{Inverse variance weighted noise estimate the respective experiment and \ccatprime{} 345 GHz channels.}
\tablenotetext{\dagger}{Five meter Three-mirror anastigmat (TMA) telescope design \citep{cmbs4collab19}.}
\end{flushleft}
\end{deluxetable*}

In this study, we consider the current survey being conducted on the South Pole Telescope (SPT) with the \sptthreeg\ camera \citep{benson14, bender18}, and the upcoming CMB experiments Simons Observatory (SO, \citealt{SO18}) and CMB-S4 \citep{cmbs4collab19} as our baseline surveys.
We use both the wide (\sfour) and the deep (\sfourdeep) CMB-S4 surveys \citep{cmbs4collab19}. 
For SO, we consider both the nominal \sofid{} and also its alternative \sogoal{} configurations \citep{SO18}.
Since mitigating the contamination from CIB is crucial for kSZ, we also include the high frequency (HF) channels from \sptfour{} \citep{anderson22} for \sptthreeg{} and Fred Young Submillimeter Telescope (\ccatprime, \citealt{CCATprime23}) for SO. 
\sptfour{} is a proposed successor of \sptthreeg{} which will observe the same 1500 $\sqdeg$ footprint of 
\sptthreeg{} but with three HF bands: 220, 285, and 345 GHz \citep{anderson22}.

The experimental configurations (beams and noise levels for each band) of the surveys are given in Table~\ref{tab_exp_specs}.
As indicated in the table, we make inverse variance weighted combinations of the overlapping bands when the HF channels are included. 
The specifications for modeling the atmospheric noise \citep{dibert22} for SPT and CMB-S4 can be found in Table~\ref{tab_exp_atm_noise_cmbs4_spt}. 
For SO, we use the values quoted in \citet{SO18} similar to the setup described in \citet{raghunathan22b}.
Throughout this study, we ignore data from channels below 90 GHz as they are primarily used to reduce contamination from the galactic synchrotron signals on large-scales which have negligible impact on both kSZ and CMB lensing. 

\begin{deluxetable}{| c | c | c | c |}
\tabletypesize{\small}
\tablecaption{Atmospheric $1/f$ noise specifications ($\ell_{\rm knee}$, $\alpha_{\rm knee}$) for  \spt{} and  \cmbsfour{} experiments.}
\label{tab_exp_atm_noise_cmbs4_spt}
\tablehead{
Band [GHz] & \spt & \sfourdeep & \sfour
}
\startdata
\hline
90 & 1200, 3.0 & 1200, 4.2 & 1932, 3.5 \\\hline
150 & 2200, 4.0 & 1900, 4.1 & 3917, 3.5 \\\hline
220 & 2100, 3.9 & 2100, 4.1 & 6740, 3.5 \\\hline
285 & 2100, 3.9 &  2100, 3.9 & 6792, 3.5 \\\hline
345 & 2600, 3.9 &  - & - \\
\enddata
\end{deluxetable}

\subsection{Simulations}
\label{sec_sims}
Thanks to the advancement in the field of computational astrophysics, we have several correlated multi-component simulations \citep{sehgal10, stein20, omori22} that are crucial to test and understand the biases in the measurements from future CMB surveys.
We primarily use the \agora{} simulation released recently by \citet{omori22} to quantify the systematics introduced by CIB, tSZ, and radio signals in both kSZ and lensing measurements. 
We do not note significant differences when replacing the frequency dependent ILC weights computed using \agora{} with a foreground model based on SPT measurements \citepalias{reichardt21}. 
More details about this validation can be found in Appendix~\ref{appendix_validation_spt_fg_model}. 
As a further check, we replace the \agora{} with \websky{} \citep{stein20} simulations in Appendix~\ref{appendix_validation_websky}. 
We limit these checks to the \sfour{} survey. 

\subsubsection{\textsc{Agora}}
We provide a brief description of \agora{} simulations here and refer the interested readers to the original paper \citep{omori22} for more details.
\agora{} is a set of correlated extragalactic maps generated based on the \mbox{MultiDark Planck 2} (MDPL2; \citealt{klypin2016}) simulation using its dark matter particles and halo catalog. The simulation consists of high resolution maps of lensed CMB, tSZ, kSZ, CIB and radio sources as well as lensing maps. The tSZ map is generated by pasting halo profiles that are fit to the \textsc{BAHAMAS} hydrodynamical simulation \citep{bahamas, mead2020}  onto the MDPL2 haloes.\footnote{Halo catalogs are publicly available \url{http://halos.as.arizona.edu/simulations/MDPL2/hlists/}} The CIB and radio catalogs are based on the MDPL2  UniverseMachine catalogs \citep{behroozi2019}.\footnote{Available at \url{http://halos.as.arizona.edu/UniverseMachine/DR1/MDPL2_SFR/}.} 

\agora{} has been verified to not only reproduce auto-/cross- frequency spectra from data but also component separated maps such as Compton-$y$ maps \citep{omori22}. In this study, we run measurements on the uncalibrated\footnote{Uncalibrated here means that the simulated maps have been convolved with the experiment-dependent frequency band pass functions, but have not been adjusted to match with the measured power spectra, which is typically a 5-10\% offset.} 
frequency maps.

\subsubsection{Sky components and masking}
\label{sec_components_and_masking}

For the kSZ study, we use the CIB, radio and tSZ signals from both \agora{} and \websky{} simulations. 
We identify the locations of dusty and radio point sources with flux $S_{\rm 150} \ge 3\ \mjy$ which roughly corresponds to $\ge 10\sigma$ detection limit from the surveys considered in this work. Note that is a conservative choice and we can reduce the masking threshold down to $5\sigma$ which corresponds to $S_{\rm 150} \sim 1-2\ \mjy$ \citep{cmbs4collab19, SO18}. 
We only mask a single pixel for the point sources. 
We also mask the location of clusters with mass $\mvir \ge \clustermask\ \msol$ which roughly corresponds to the $10\sigma$ detection limit \citep{raghunathan22, raghunathan22b}.
For the cluster mask, we do not modify the masking threshold as a function of redshift. 
However, the size of the mask changes based on the redshift of the halo and we define the masking radius to be $2\thetavir$. 

For CMB lensing, besides the astrophysical CIB, radio, and SZ signals, we also use the lensed CMB maps with the associated convergence field. The point source masking threshold used for CMB lensing is higher ($S_{\rm 150} \ge 6\ \mjy$) than the one adopted for kSZ. This choice is primarily to reduce the mean-field lensing bias in the reconstructed lensing map. 
The cluster masking threshold is $10\sigma$.

\subsection{Internal Linear Combination}
\label{sec_ilc}

\begin{figure}
\centering
\ifdefined\ApJsubmit
\includegraphics[width=0.45\textwidth, keepaspectratio]{ilc_residuals_mv.pdf}
\else
\includegraphics[width=0.45\textwidth, keepaspectratio]{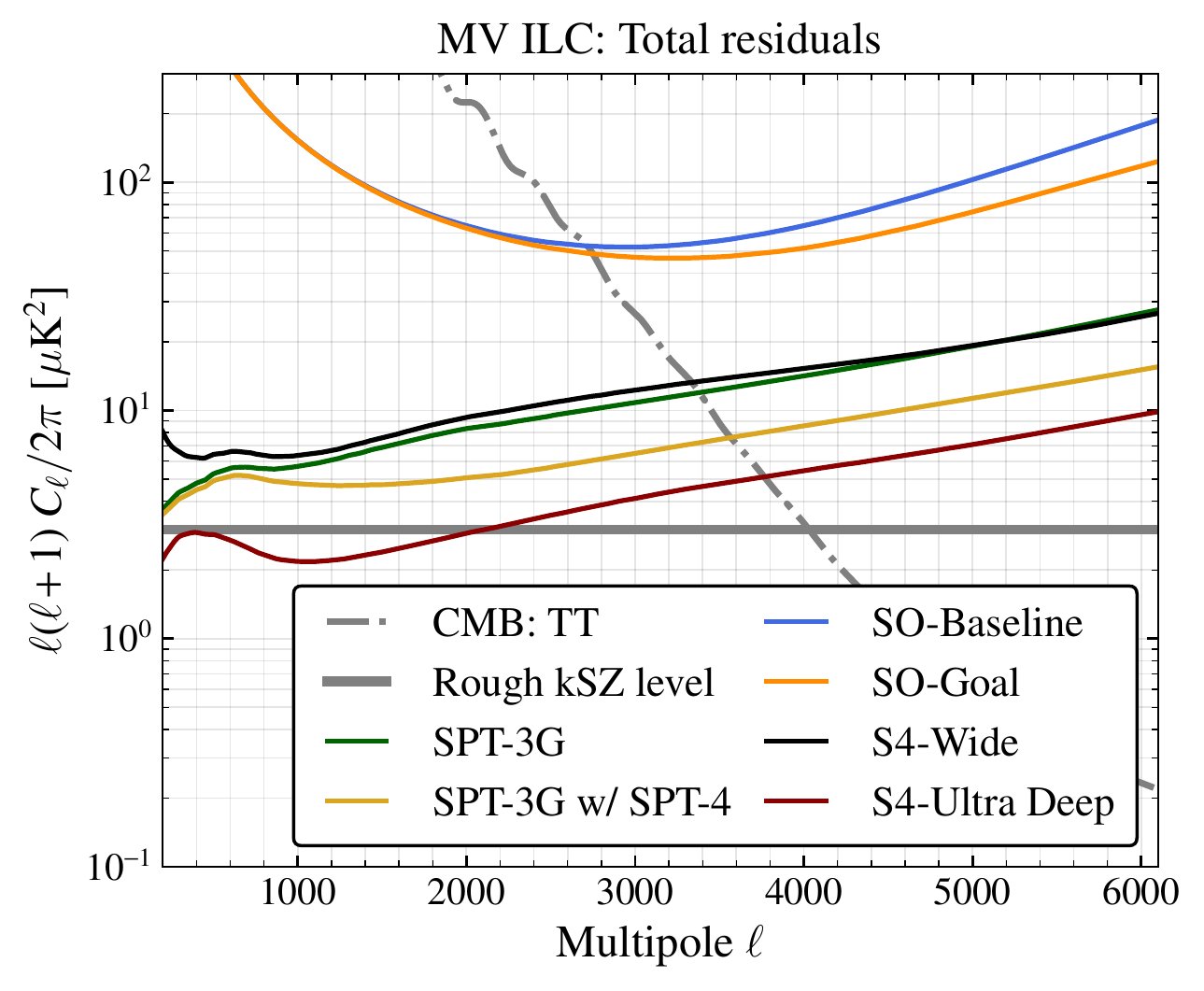}
\fi
\caption{The total ILC residuals in the MV CMB map. The solid gray line shows the primary CMB TT power spectrum while the dash-dotted gray line corresponds to a rough kSZ level of $\dl = \kszleveldl \uk^{2}$. CMB dominates on large scales ($\ell \lesssim 4000$) while the residual noise and foregrounds are higher than kSZ on small scales.
 We note that the kSZ \snr{} for individual modes is always lower than one for all experiments.}
\label{fig_mvilc_residuals}
\end{figure}

\begin{figure*}
\centering
\ifdefined\ApJsubmit
\includegraphics[width=0.9\textwidth, keepaspectratio]{ilc_cib_tsz_residuals_mv.pdf}
\else
\includegraphics[width=0.9\textwidth, keepaspectratio]{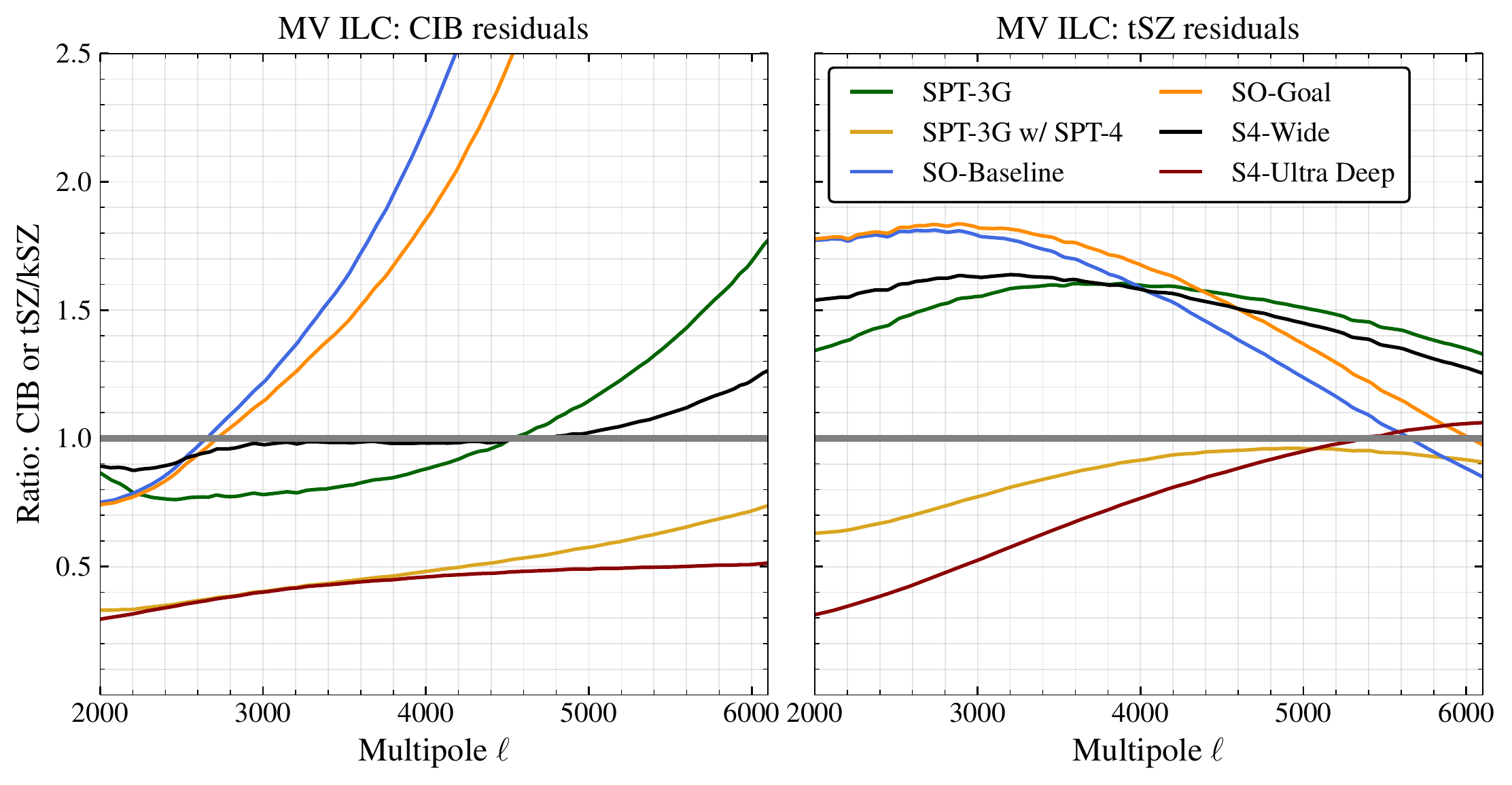}
\fi
\caption{The residual CIB (left) and tSZ (right) signals compared to the expected kSZ level of $\dl = \kszleveldl \uk^{2}$ in the MV CMB/kSZ maps from different experiments considered in this work.}
\label{fig_cib_tsz_mvilc_residuals}
\end{figure*}

We combine data from different frequency channels ${\rm N}_{\rm ch}$ using the ILC technique in the harmonic space as 
\begin{eqnarray}
S_{\ell} = \sum_{i=1}^{\rm N_{\rm ch}} w_{\ell}^{i} M_{\ell}^{i},
\end{eqnarray}
where $S$ corresponds to the desired sky signal, which in this case is the sum of CMB and kSZ signals.
The multipole dependent weights $w_{\ell}$ for each frequency channel $i$ given in 
Eq.(\ref{eq_ilc_weights})
are computed to minimize the total variance from experimental noise and foregrounds to obtain the minimum-variance (MV) signal map. 
The weights can also be tuned to produce a minimum-variance map along with an additional constraint of nulling the contribution to a specific frequency response. 
This is the cILC approach.
The weights are constructed as
\begin{eqnarray}
w_{\ell}^{\rm cILC} & = &  \clinv \mathcal{F} \left(\mathcal{F}^{\dagger} \clinv \mathcal{F} \right)^{-1} N,
\label{eq_ilc_weights}
\end{eqnarray}
where the matrix ${\bf C}_{\ell}$ has a dimension $\rm N_{ch} \times \rm N_{ch}$ and contains the covariance between simulated maps in multiple frequencies at a given multipole $\ell$; \mbox{$\mathcal{F} = [A_{S}\ B_{S}\ C_{S}\ ...\ Z_{S}]_{{\rm N}_{\rm ch} \times S}$} contains the frequency response vector of the signal component $S$, either the desired signal $A_{S}$ with dimension \mbox{$\rm N_{\rm ch} \times 1$} or the undesired sky components $[B_{S}\ C_{S}\ ...\ Z_{S}]$ that are being nulled using the cILC \citep{remazeilles11} technique which is specified using \mbox{$N = [1\ 0\ 0\ ..\ 0]_{S \times 1}$}. 

For a standard MV ILC, Eq.(\ref{eq_ilc_weights}) simplifies to \mbox{$w_{\ell}^{\rm MV} =  \dfrac{\clinv A_{S}}{A_{S}^{\dagger} \clinv A_{S}}$}.
In Fig.~\ref{fig_mvilc_residuals}, we show the total (noise and foregrounds) ILC residuals for the standard MV ILC estimator for different experiments considered in this work. 
The foreground signals are from \agora{} simulations. 
The primary CMB (gray solid curve) is much larger than the kSZ (gray dash-dotted curve) on large scales ($\ell \lesssim 4000$) while the small scales are dominated by instrumental noise and foregrounds.

The instrument noise is generally easy to model. 
The contribution from radio point sources in desired frequency band $\nu$ below the masking threshold can also be modeled relatively easily assuming an underlying source distribution $dN/dS$ as we show later.
On the other hand, the residual CIB and tSZ signals are harder to model. 
The contributions from CIB and tSZ in the MV CMB map relative to the kSZ signal $\dl = \kszleveldl \uk^{2}$ are presented in Fig.~\ref{fig_cib_tsz_mvilc_residuals} in the left and right panels respectively. 
As it is evident from the figure, the residual CIB and tSZ signals are non negligible for \sptthreeg{} (green), \sfour{} (black), \sofid{} (blue), and \sogoal{} (orange). 
The CIB residuals are $\sim \times 2$ lower than kSZ for \sptfour{} (yellow) and \sfourdeep{} (red) although the tSZ residuals become higher than kSZ for $\ell \ge 4000$. 
This suggests that using the MV ILC map for kSZ measurements could lead to significant biases depending on the uncertainties in modeling the CIB and tSZ signals (see Appendix~\ref{appendix_bias_cibtsz}). 

\subsection{Cross-ILC}
\label{sec_cross_ilc}

\begin{figure}
\centering
\ifdefined\ApJsubmit
\includegraphics[width=0.45
\textwidth, keepaspectratio]{ilc_cib_plus_tsz_residuals_s4_wide.pdf}
\else
\includegraphics[width=0.45\textwidth, keepaspectratio]{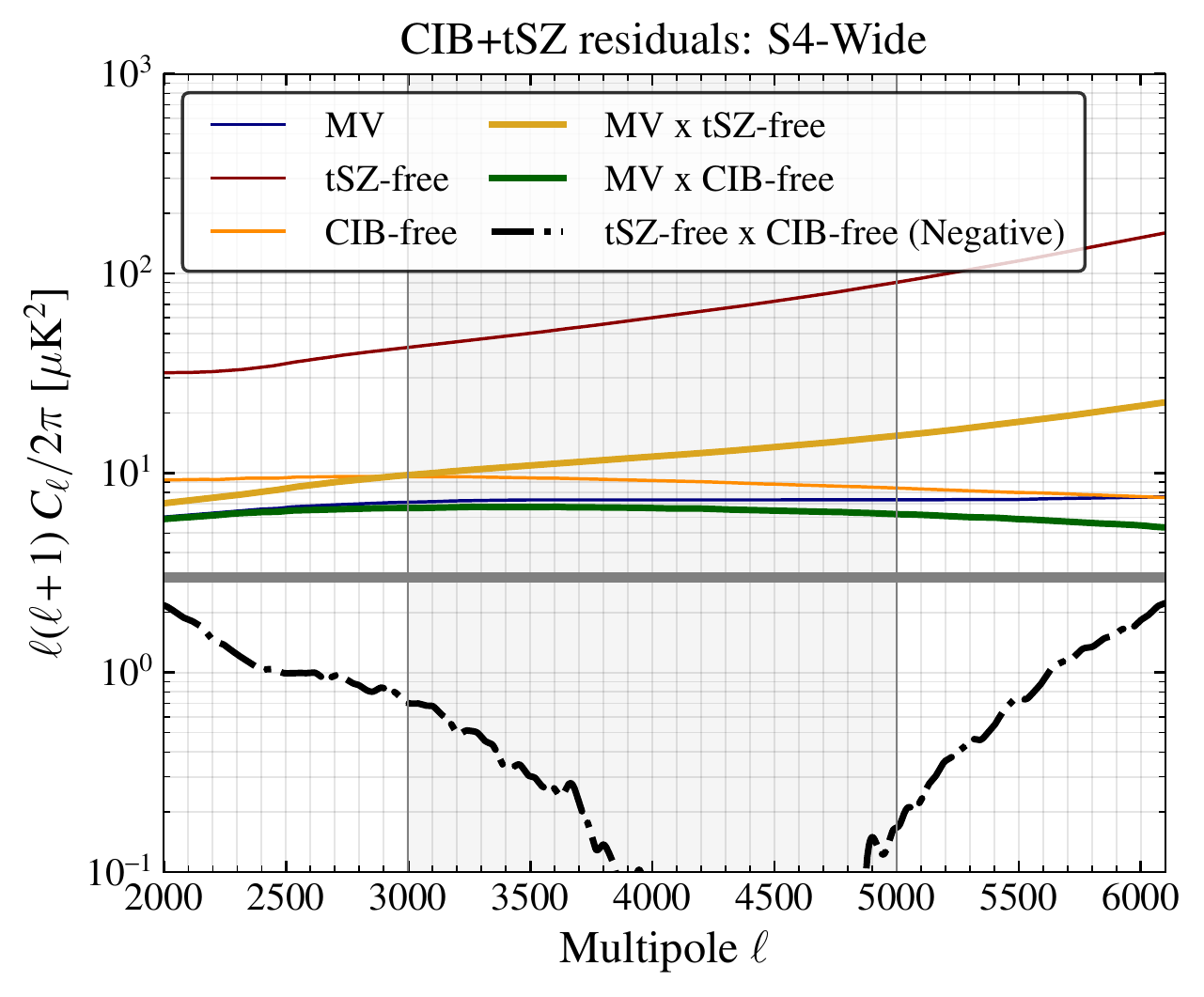}
\fi
\caption{The total CIB and tSZ residuals expected in the \sfour{} survey for different ILC combinations: Thin lines are for auto-spectra and thick lines are for cross-ILC. 
We note that the residuals in cross-ILC combination (\tszfree{} $\times$ \cibfree{} in black) is much lower than kSZ ($\dl = \kszleveldl \mu {\rm K}^{2}$) and hence can be a compelling means of measuring the kSZ power spectrum with future CMB surveys. 
}
\label{fig_cib_plus_tsz_allilc_residuals}
\end{figure}


As discussed above, it is possible to reduce the contamination from CIB/tSZ using the cILC method in Eq.(\ref{eq_ilc_weights}).
This can be achieved in multiple ways. 
For example, 
\begin{itemize}
\item[(A)]{Creating a single ILC map by jointly nulling tSZ and CIB using $B_{S}$ and $C_{S}$ in Eq.(\ref{eq_ilc_weights}) where $B_{S}$ and $C_{S}$ are the frequency response of tSZ and CIB SED being nulled. In this case, we will take the auto-spectrum of this cILC map.
}
\item[(B)]{Creating two cILC maps. The first map is a tSZ-nulled map such that $B_{S}$ is the frequency response of the tSZ signal. The second map is a CIB-nulled map and here $B_{S}, C_{S}$ correspond to the CIB SED(s) being nulled. In this case, we will then take the cross-spectrum of the two cILC maps.}
\item[(C)]{This is similar to the second method but the CIB SED(s) for nulling are chosen such that the total CIB+tSZ residuals are lower in the cross-spectrum. {\bf \emph{This is the approach we adopt in this work.}}}
\end{itemize}

\noindent
{\bf Approach (A)}: If CIB signal is made up of a single population (i.e:) if the frequency response can be described by a single SED, then the first approach is the optimal for mitigating CIB.
However, the CIB signal is made up of multiple galaxy populations at different redshifts and nulling the tSZ significantly enhances the contamination from CIB populations that are not included in Eq.(\ref{eq_ilc_weights}). 
Moreover, this approach increases the noise significantly by $\times2-\times3$ compared to the MV ILC. 

\noindent
{\bf Approach (B)}: The noise in approach (B) is lower than (A). However, it does not take the cross-correlation CIB$\times$tSZ into account and hence not ideal. 

\noindent
{\bf Approach (C)}: The third approach (C) is similar to (B), but we optimise the weights for CIB nulling in a manner that reduces the total CIB+tSZ residuals in the CMB map. Thus, it also takes the cross-correlation CIB$\times$tSZ into account. 
We describe the steps to obtain weights for tSZ and CIB nulling below.

In Fig.~\ref{fig_cib_plus_tsz_allilc_residuals}, we compare the total CIB+tSZ residuals expected in \sfour{} for different ILC techniques. 
Thin lines are the auto-spectra of the ILC maps: blue is for the MV, red is for \tszfree, and orange is for a multi-component CIB-nulled ILC map. 
Thick lines are for the cross-ILC technique obtained as the cross-spectrum of two different ILC maps. 
From the figure, we note that the cross-ILC spectrum obtained from \tszfree{} and \cibfree{} maps shown in black returns the best total CIB+tSZ residual and is almost an order of magnitude lower than kSZ for $\ell \in [3500, 5000]$. 
As a result, we use this cross-ILC combination as the baseline in the rest of the paper. 
We provide further details in \S\ref{sec_total_cib_tsz_residuals}.
\\~\\
\noindent{\bf \emph{Weights for tSZ and CIB nulling:}} The weights for \tszfree{} map are obtained by setting \mbox{$B_{s} = [-4.36, -2.61, -0.1, 2.94, 5.72, 8.79, 29.9]$}, which correspond to the frequency response of the tSZ signal in \mbox{$\nu = [90, 150, 220, 285, 345, 410, 850]$} GHz bands, in Eq.~\ref{eq_ilc_weights}. 
The CIB signal is generally modeled as a modified blackbody as $\eta_{\nu} = \nu^{\beta_{\rm CIB}} B_{\nu}(T_{\rm CIB})$ where $T_{\rm CIB}$ is the temperature, $\beta_{\rm CIB}$ is the emissivity index, and $B_{\nu}(T)$ is the \planck{} function. 
Getting the frequency response of the CIB signal is slightly tricky as the CIB is made up of multiple populations of DSFGs with different values of $T_{\rm CIB}$ and $\beta_{\rm CIB}$. 
As a result, we do not adopt the same values of $T_{\rm CIB}$ and $\beta_{\rm CIB}$ for all experiments. 
Instead, we perform a blind four parameter ($T^{1}_{\rm CIB}, \beta^{1}_{\rm CIB}, T^{2}_{\rm CIB}, \beta^{2}_{\rm CIB}$) grid search to obtain a two component SED for CIB-nulling and the frequency response of these two SEDs are plugged into $B_{s}$ and $C_{s}$ of Eq.~\ref{eq_ilc_weights}. 
The SEDs that return the lowest level of CIB+tSZ residual in the multipole window $\ell \in [3000, 5000]$ (gray band in Fig.~\ref{fig_cib_plus_tsz_allilc_residuals} and Fig.~\ref{fig_cib_plus_tsz_residuals_crossilc_all_experiments}) are chosen. 
We use this range for kSZ as the sample variance from CMB fully swamps the kSZ detection at $\ell \le 3000$. 
For lensing, we use a slightly different range $\ell \in [2000, 5000]$. 
We also checked the results with a single SED or a three-component SED model and find that the two-component SED works the best in terms of both $\snr$ and foreground residuals. 
Note that the above weights are calculated using \agora{} foreground model but, as mentioned earlier, we validate this assumption using a foreground model based on SPT measurements \citepalias{reichardt21} and Websky simulations \citep{stein20} in Appendix~\ref{appendix_validation_fg_model}.

\subsubsection{Fisher formalism}
\label{sec_fisher_formalism}

\newcommand{\logAs}{{\rm ln}(10^{10}A_{s})}
\begin{deluxetable}{| c | c | c |}
\tabletypesize{\small}
\def\arraystretch{1.1}
\tablecaption{Fiducial values of the parameters and priors used in this work. All the applied priors are Gaussian with widths given below in the table.}
\label{tab_parameters_priors}
\tablehead{
Parameter & Fiducial & Prior
}
\startdata
\hline
\multicolumn{3}{l}{\it Cosmological parameters:}\\\hline
Amplitude of scalar fluctuations $\logAs$ & 3.044 & \multirow{6}{*}{-}\\\cline{1-2}
Dark matter density $\omchsq$ & 0.1200 & \\\cline{1-2}
Baryon density $\ombhsq$ & 0.02237 & \\\cline{1-2}
Scalar spectral index $n_{s}$ & 0.9649 & \\\cline{1-2}
Angular size of sound horizon & \multirow{2}{*}{1.04092} & \\
at recombination $100\theta_{\ast}$ & & \\\hline
\multirow{2}{*}{Reionization optical depth $\taure$} & \multirow{2}{*}{0.0544} & 0.007\tablenotemark{$^{a}$} \\
 & & 0.002\tablenotemark{{$^{b}$}} \\\hline\hline
\multicolumn{3}{l}{\it Foreground parameters:}\\\hline
Residual CIB+tSZ $\acibtsz$ & 1 & 0.1\\\hline
Mean spectral index of sources $\alpharad$ & -0.76 & 0.1 \\\hline
Scatter in source spectral indices $\alpharadsigma$ & 0.2 & 0.3 \\\hline\hline
\multicolumn{3}{l}{\it Total kSZ power spectrum:}\\\hline
Amplitude $\aksz$ & 1 & - \\\hline\hline
\multicolumn{3}{l}{\it Homogeneous kSZ power spectrum:}\\\hline
Amplitude $\akszhomo$ & 1 & 0.1 \\\hline
Spectral tilt $\alphakszhomo$ & 0 & 0.1 \\\hline\hline
\multicolumn{3}{l}{\it Reionization kSZ:}\\\hline
\multirow{2}{*}{Mid-point of reionization $\zmid$} & \multirow{2}{*}{7.69} & 1.43\tablenotemark{$^{a\dagger}$} \\
 & & 0.41\tablenotemark{$^{b\dagger}$} \\\hline
Duration of reionization $\zdur$ & 4 & - \\\hline
\enddata
\tablenotetext{a}{\planck-like $\taure$ prior.}
\tablenotetext{b}{\litebird-like $\taure$ prior.}
\tablenotetext{\dagger}{\planck-like or \litebird-like $\taure$ prior translated to prior on $\zmid$.}
\end{deluxetable}
We use a Fisher formalism to forecast the expected \snr{} for the total kSZ power spectrum. 
For this purpose, we use the cross-ILC combination for all experiments and also compare it with the \snr{} obtained from the MV ILC. 
Although MV ILC is expected to return the best \snr, we note from Fig.~\ref{fig_cib_tsz_mvilc_residuals} and Fig.~\ref{fig_cib_plus_tsz_allilc_residuals}, the total ILC residuals receive significant contribution from CIB and tSZ; and hence will be prone to foreground-induced biases. 

For kSZ forecasts, we consider two approaches. 
In the first case, we compute the total kSZ \snr{} and for this we use a ten parameter model of which six are the standard $\lcdm$ parameters \mbox{$\theta \in [A_{s},\ h,\ n_{s},\ \omchsq,\ \ombhsq,\ \taure]$} and the additional parameters are: $\aksz$ which quantifies the amplitude of the kSZ signal $\dl = \aksz \times \kszleveldl \uk^{2}$; $\acibtsz$ which represents the amplitude of the total CIB and tSZ residuals; $\alpharad$ and $\alpharadsigma$ which represent the mean spectral index and the scatter when modeling the radio residuals. 

In the second approach, we replace the total kSZ amplitude $\aksz$ with parameters that govern the physics of reionization namely: $\zmid$ which is the mid-point of reionization and $\zdur$ which is the duration of reionization. 
Since, it is hard to disentangle between the two kSZ components (homogeneous and patchy kSZ) from the total kSZ power spectrum, we include two more parameters along with suitable priors $\akszhomo$ and $\alphakszhomo$ that represent the amplitude and the slope of the homogeneous kSZ power spectrum. Thus in the second case, we fit for 13 parameters.

We use CMB TT/EE/TE power spectra since EE/TE will dominate the $\lcdm$ parameter constraints from future CMB surveys \citep{calabrese14, galli14}. 
For EE/TE, we use data in the multipole range $\ell \in [30, 5000]$. 
For TT, we use a slightly different strategy. 
To better predict the radio residuals, we use $\ell_{\rm max}^{\rm TT} = 5000$. 
However, following other similar works in the literature \citep[][]{calabrese14, SO18, alvarez20}, we set different values of $\ellmaxksz \in [3000, 3500, 4000, 4500]$ and 
do not consider modes above $\ellmaxksz$ for kSZ to account for the small-scale foreground residuals. We do this by setting the derivatives of the kSZ power spectrum $\dfrac{\partial kSZ}{\partial \aksz} = 0$ above $\ellmaxksz$. 
The fiducial values of the parameters and the priors are listed in Table~\ref{tab_parameters_priors}.

\section{Results and Discussion}
\label{sec_results}

\subsection{Kinematic SZ}
\label{sec_results_ksz}

\begin{figure*}
\centering
\ifdefined\ApJsubmit
\includegraphics[width=0.9\textwidth, keepaspectratio]{ilc_cib_plus_tsz_residuals_crossilc.pdf}
\else
\includegraphics[width=0.9\textwidth,keepaspectratio]{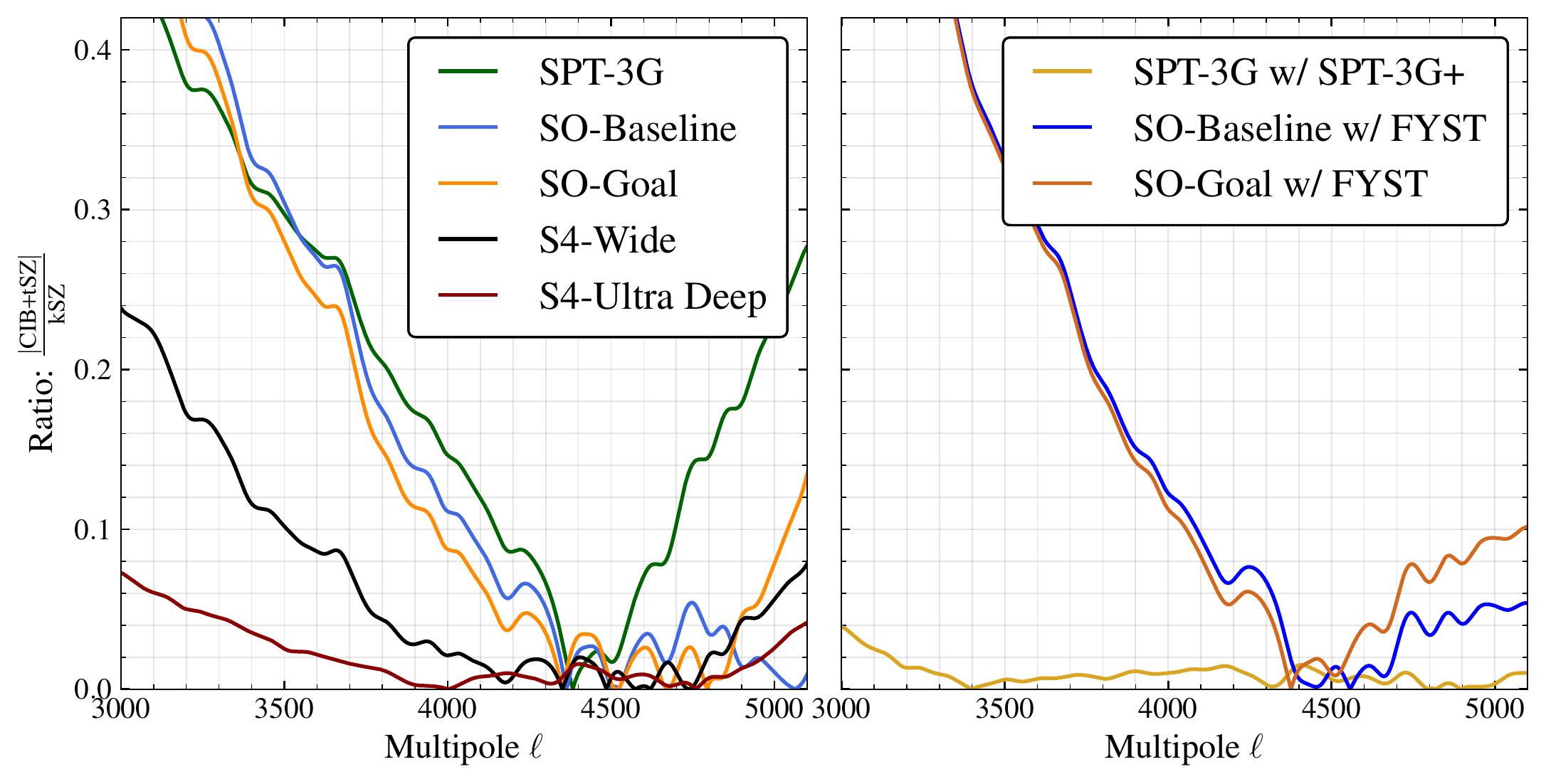}
\fi
\caption{The ratio of the total CIB+tSZ residual compared to the expected level of kSZ ($\dl = \kszleveldl \mu {\rm K}^{2}$) for the cross-ILC combination \tszfree{} $\times$ \cibfree{} from different CMB surveys. In the left, we show the residuals for the baseline experiments while the right panel is after the inclusion of HF bands from \sptfour{} or \ccatprime. 
}
\label{fig_cib_plus_tsz_residuals_crossilc_all_experiments}
\end{figure*}

In this section, we first show the improvement in CIB and tSZ residual levels using the cross-ILC technique for all the experiments. Besides the CIB+tSZ residuals, the CMB maps also contain the residual signals from instrumental noise and radio galaxies. We handle these next. This is followed by the calculation of the expected kSZ SNR using the Fisher formalism. 

\subsubsection{Total CIB+tSZ residuals}
\label{sec_total_cib_tsz_residuals}
In Fig.~\ref{fig_cib_plus_tsz_residuals_crossilc_all_experiments}, we present the ratio of the total CIB+tSZ residuals over the expected kSZ signal ($\dl = \kszleveldl \uksq$) for the cross-ILC combination \tszfree{} $\times$ \cibfree{} for different experiments. 
The left panel contains the residuals for the baseline experiments. 
As it is evident, the total CIB+tSZ residuals are always lower than kSZ for the cross-ILC combination. 
To be specific, the residuals are roughly $\times2$ ($\times3$) lower than kSZ in $\ell \in [3000, 5000]$ ($\ell \in [3500, 5000]$) for all experiments. 
For CMB-S4, the residuals are $\le \times 10$ lower than kSZ in the range $\ell \in [3500, 5000]$.
By comparing Fig.~\ref{fig_cib_plus_tsz_residuals_crossilc_all_experiments} to Fig.~\ref{fig_cib_tsz_mvilc_residuals}, we can deduce that the kSZ power spectrum can be measured by all the experiments using the cross-ILC combination much more robustly compared to the MV ILC estimator.
\\~\\
\noindent
{\bf \emph{Inclusion of HF bands:}} In the right panel of  Fig.~\ref{fig_cib_plus_tsz_residuals_crossilc_all_experiments}, we assess the performance after the inclusion of HF bands to the baseline experiments: \sptfour{} to \sptthreeg{} and \ccatprime{} to SO. 
Since the residuals are already small for \sfour{} and \sfourdeep, we do not include \ccatprime{} for CMB-S4.
For SPT, we find a significant improvement in the reduction of CIB+tSZ residuals after including information from \sptfour. 
The noise, although not shown here, also reduces for \sptthreeg. 
For SO, however, including \ccatprime{} does not improve the residuals. 
In fact, from the right panel of Fig.~\ref{fig_cib_plus_tsz_residuals_crossilc_all_experiments}, we note that the residuals have slightly increased. 
On the other hand, addition of \ccatprime{} does reduce the overall noise in the ILC maps by $\times1.5$ to $\times2$ compared to SO-only at $\ell \in [4000, 5000]$. 
This behavior is not surprising and it is because of the higher noise levels of SO and \ccatprime{} compared to \sptthreeg{} and \sptfour. 
In the case of the former, the ILC algorithm optimises the weights to reduce the overall noise primarily while for the latter, the weights are optimised to reduce the foreground residuals.
The noise reduction ultimately helps in improving the final kSZ \snr{} for SO as shown in Table~\ref{tab_total_ksz_snr}. 

We note here that the total CIB+tSZ residuals can be lowered further for \mbox{SO + \ccatprime} at the expense of a higher noise by introducing a scaling term for noise or CIB in the covariance matrix $\clcov$ in Eq.(\ref{eq_ilc_weights}) used for computing the ILC weights as prescribed by \citet{bleem21}. 
While this approach is not optimal for \snr, it does help in reducing the residuals. 
Following this idea, when we scaling the noise levels of \ccatprime's HF channels ($\nu \in [345, 410, 857]$) by $\times5$, we find the CIB+tSZ residuals at $\ell \in [4000, 5000]$ to go down by $\times1.5$ for \sogoal.



\subsubsection{Handling radio residuals}
\label{sec_radio_residuals}

Given that the power spectrum of the radio galaxies follow a Poisson distribution \citep{gonzalezneuvo05, lagache19}, they can modeled easily compared to other foreground signals. 
In this work, we use a point source masking threshold of $S_{150}^{\rm max}$ = 3 mJy. This is a conservative choice and the masking threshold can be lowered further for both current and future experiments. 
We model the power from sources below the masking threshold by assuming an underlying source distribution $dN/dS$ along with a mean spectral index $\alpharad$ and a scatter $\alpharadsigma$ as 

\begin{eqnarray}
\label{eq_radio_point_source_power} 
C_{\ell_{{\nu_{1} \nu_{2}}}}^{\rm radio} =  \bigintsss_{\alpharad^{\rm min}}^{\alpharad^{\rm max}} d\alpha \bigintsss_{0}^{S_{150}^{\rm max}} dS_{150}\ S_{150}^{2}\\
  \frac{dN}{dS_{150}}\ \left( \frac{\nu_{0}^{2}}{\nu_{1} \nu_{2}} \right)^\alpha \mathcal{N} \left[ \alpharad | \bar{\alpha}_{\rm rad}, \alpharadsigma \right]\notag,
\end{eqnarray}
where the normal distribution $\mathcal{N} \left( \mu, \sigma\right)$ is used to parameterize the probability density function of source spectral indices.

\begin{figure}
\centering
\ifdefined\ApJsubmit
\includegraphics[width=0.45\textwidth, keepaspectratio]{ilc_radio_residuals_s4_wide_tszfreexcibfree.pdf}
\else
\includegraphics[width=0.45\textwidth, keepaspectratio]{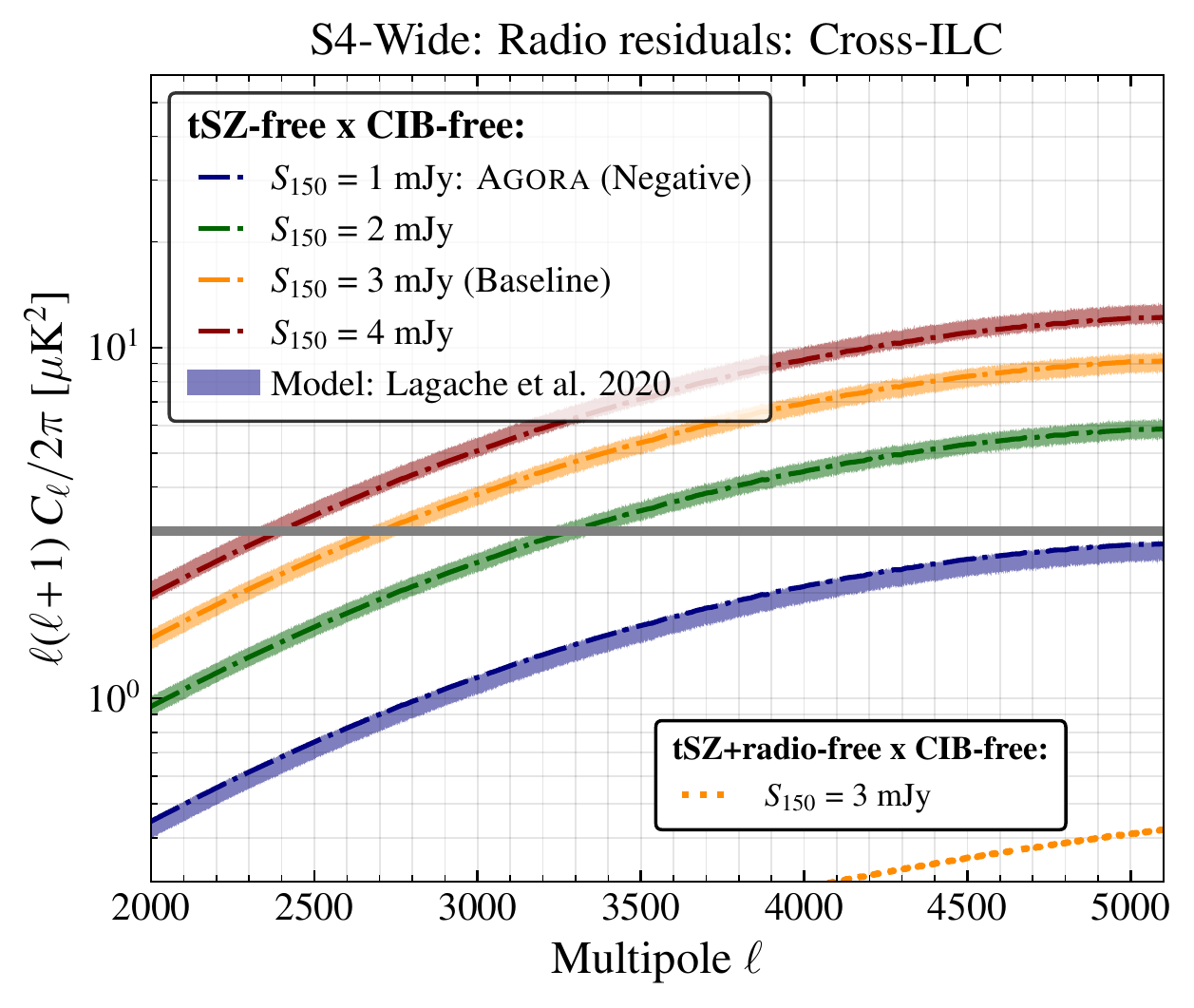}
\fi
\caption{Radio residuals using \agora{} expected from \sfour{} for the masking threshold $S_{150} = 3\ {\rm mJy}$ (yellow curve) used in this work. They are negative for the cross-ILC \tszfree{} $\times$ \cibfree{} combination. We also show the analytical modeling of the radio residuals using \citet{lagache19} in the yellow band. 
The band quantifies the residual power assuming a scatter in the radio spectral index $\alpharadsigma \in [0, 0.35]$. 
For reference, we also show the residuals and the models for multiple masking thresholds: $S_{150} = 1\ {\rm mJy}$ in blue, $S_{150} = 2\ {\rm mJy}$ in green, and $S_{150} = 4\ {\rm mJy}$ in red. 
The yellow dotted curve near the bottom right is the residual expected when a radio spectrum $\alpharad = -0.76$ is nulled along with tSZ.}
\label{fig_radio_residual_modelling}
\end{figure}

The residual radio signal in the ILC combination can now be calculated analytically using Eq.(\ref{eq_radio_ilc_residuals}) as
\begin{eqnarray}
C_{\ell_{\rm ILC}}^{\rm radio} & = & w_{\ell_{A}} \clcov^{\rm radio}\ w_{\ell_{B}}^{\dagger}
\label{eq_radio_ilc_residuals}
\end{eqnarray}
where $\clcov^{\rm radio}$ is the ${\rm N}_{\rm ch} \times {\rm N}_{\rm ch}$ covariance matrix containing the auto- and cross-power spectra between different frequency bands; and $w_{\ell_{A}}$ and $w_{\ell_{B}}$ are the frequency dependent weights for the two ILC maps $A$ and $B$. 

In Fig.~\ref{fig_radio_residual_modelling} we show the radio residual (yellow curve) for the cross-ILC \tszfree{} $\times$ \cibfree{} combination expected in the \sfour{} survey after masking source above $S_{150} = 3\ {\rm mJy}$, our baseline masking threshold.
The radio residuals are negative for this cross-ILC combination. 
For the analytical prediction of the radio residuals, we use \citet{lagache19} radio source distribution model $dN/dS$ with $\alpharad = -0.76$ \citepalias{reichardt21}. 
The yellow band in the figure corresponds to radio residuals assuming a scatter in the spectral index $\alpharadsigma \in [0, 0.35]$. 
As can be seen from Fig.~\ref{fig_radio_residual_modelling}, the width of the band is: $D_{\ell_{\rm radio}}^{\rm sys} = [0.4, 0.8, 1.0]$ for $\ell \in [3000, 4000, 5000]$ and these are smaller than the expected kSZ level of \mbox{$\dl = \kszleveldl \uk^{2}$}. 
For reference, we also show the radio residuals and the predictions for other masking threshold: $S_{150} = 1\ {\rm mJy}$ in blue, $S_{150} = 2\ {\rm mJy}$ in green, and $S_{150} = 4\ {\rm mJy}$ in red. 
When computing the kSZ SNR in \S~\ref{sec_fisher_forecasts_for_ksz}, we account for the radio residuals using two parameters $\alpharad$ and $\alpharadsigma$. 
\\~\\
\noindent
{\bf \emph{Nulling radio along with tSZ:}}
Another approach to mitigate the radio residuals is to null the response to radio sources along with the tSZ removal in the first leg of the cross-ILC.
This can be achieved by setting the frequency responses $B_{s}, C_{s}$ in Eq.(\ref{eq_ilc_weights}): \mbox{$B_{s} = [-4.36, -2.61, -0.1, 2.94]$} is the frequency response of tSZ and \mbox{$C_{s} = [2.89, 1.0, 0.62, 0]$} in the frequency response of radio signals for $\alpha = -0.76$ in the bands \mbox{$\nu = [90, 150, 220, 285]$}. 

We show the reduction in radio power using this method in Fig.~\ref{fig_radio_residual_modelling} for our baseline masking threshold $S_{150} = 3\ {\rm mJy}$ as the yellow dotted curve (near the bottom right corner). 
Since not all the radio sources have the same spectral index $\alpharad = -0.76$, we can expect to have some amount of residual signal as can be seen in Fig.~\ref{fig_radio_residual_modelling}. 
As it is evident from the yellow dotted curve, this residual is an order of magnitude smaller than kSZ and hence negligible. 
This curve also includes a band around it with $\alpharadsigma = 0.4$ although the width of the band is small and hence not visible.
Note that including an additional constraint for nulling radio along with tSZ in cILC will increase the noise level of the resultant map compared to the \tszfree{} map. 
Hence, we present this method only as a proof of concept and do not pursue it further in this work. 

\begin{deluxetable*}{| c | c || c | c || c | c || c | c || c | c ||}
\tabletypesize{\small}
\renewcommand{\arraystretch}{1.3}
\tablecaption{Forecasted \snr{} of the total kSZ power spectrum $\dl = \kszleveldl\ \mu K^2$ for different experiments using the Fisher\tablenotemark{\text{\faGithub}} formalism. 
These are obtained after marginalizing over the $\lcdm$ parameters and $\acibtsz$. 
\planck-like prior on $\sigma(\taure)=0.007$ and a 10\% prior on $\sigma(\acibtsz)=0.1$ have been assumed. 
The radio residual modelling parameters $\alpharad$ and $\alpharadsigma$ are fixed for this table.
The associated biases in $\aksz$ due to unmodeled CIB and tSZ residuals are presented in Fig.~\ref{fig_appendix_ksz_bias_cibtsz} of Appendix~{\ref{appendix_bias_cibtsz}}.
}
\label{tab_total_ksz_snr}
\tablehead{
\multirow{3}{*}{Experiment} & \multirow{3}{*}{$f_{\rm sky}$} & \multicolumn{8}{c|}{Total kSZ SNR = $\aksz / \sigma(\aksz)$}\\
\cline{3-10}
& & \multicolumn{2}{c||}{$\ellmaxksz = 3000$} & 
\multicolumn{2}{c||}{$\ellmaxksz = 3500$} & 
\multicolumn{2}{c||}{$\ellmaxksz = 4000$} & 
\multicolumn{2}{c||}{$\ellmaxksz = 4500$} \\
\cline{3-10}
& & MV & Cross-ILC
& MV & Cross-ILC
& MV & Cross-ILC
& MV & Cross-ILC
}
\startdata
\hline\hline
\sptthreeg{} & \multirow{2}{*}{\fskydeepfull} & 10.33 & 8.39 & 19.73 & 12.94 & 27.68 & 16.55 & 27.09 & 19.11\\\cline{1-1}\cline{3-10}
+ \sptfour{} & & 11.51 & 10.90 & 23.61 & 20.23 & 35.76 & 29.20 & 37.20 & 35.77\\\hline\hline
\sofid{} & \multirow{2}{*}{\fskyso} & 15.26 & 5.13 & 24.64 & 7.86 & 30.06 & 10.71 & 25.80 & 13.13\\\cline{1-1}\cline{3-10}
+\ccatprime{} & & 15.36 & 7.30 & 24.90 & 11.44 & 30.42 & 15.82 & 26.18 & 19.60\\\hline\hline
\sogoal{} & \multirow{2}{*}{\fskyso} & 16.12 & 5.45 & 27.05 & 8.51 & 35.54 & 11.88 & 32.81 & 14.90\\\cline{1-1}\cline{3-10}
+\ccatprime{} & & 16.21 & 7.54 & 27.33 & 12.04 & 35.98 & 16.98 & 33.29 & 21.37\\\hline\hline
\sfour{} & \fskywideclean \tablenotemark{$\dagger$} & 41.68 & 32.03 & 78.45 & 48.55 & 109.33 & 67.48 & 107.79 & 83.68\\\hline
\sfourdeep{} & \fskydeepfull & 12.87 & 12.37 & 28.45 & 25.77 & 45.41 & 41.50 & 48.48 & 54.11\\\hline\hline
\enddata
\tablenotetext{\text{\faGithub}} {Jupyter notebooks  \href{https://github.com/sriniraghunathan/cross_ilc_methods_paper/blob/main/s1_get_fisher_matrix.ipynb}{\texttt{s1\_get\_fisher\_matrix.ipynb}} and \href{https://github.com/sriniraghunathan/cross_ilc_methods_paper/blob/main/s2_analyse_fisher_matrix.ipynb}{\texttt{s2\_analyse\_fisher\_matrix.ipynb}} used for forecasting are publicly available.}
\tablenotetext{\dagger}{Although \sfour{} will map \fskywidefull{} of the sky, we exclude the regions that are significantly contaminated by galactic foregrounds. See Fig. 1 of \citet{raghunathan22}.} 
\end{deluxetable*}

\subsubsection{Bandpower errors}
\label{sec_bandpower_errors}
Besides the tSZ+CIB residuals, the CMB maps contain residual noise which also contributes to the overall variance. 
The noise can be modeled easily and the noise residuals $N_{\ell_{\rm ILC}}$ in the ILC maps can be computed by replacing $\clcov^{\rm radio}$ in Eq.(\ref{eq_radio_ilc_residuals}) with ${\bf N}_{\ell}$, which is a ${\rm N}_{\rm ch} \times {\rm N}_{\rm ch}$ covariance matrix containing the auto- and cross-noise spectra between different frequency bands. 
${\bf N}_{\ell}$ is generally diagonal but given that current and future CMB surveys use multi-chroic pixels, the atmospheric noise will be correlated between bands that share the same pixel on the focal plane. 
In this work, we assume that the atmospheric noise is correlated at 90\% level between the adjacent bands \citep{SO18, cmbs4collab19}. 
Tweaking the correlation level does not affect any of our results. 
For the case of white noise, ${\bf N}_{\ell}$ is diagonal. 

With the foreground and noise residuals in hand, we can now calculate the bandpower errors as Eq.(\ref{eq_knox_errors} \citealt{knox95})
\begin{eqnarray}
\Delta \hat{C}_{\ell} & = & \sqrt{ \dfrac{2}{(2\ell + 1) f_{\rm sky} \Delta_{\ell}}}\hat{C}_{\ell}
\label{eq_knox_errors}
\end{eqnarray}
where
\begin{eqnarray}
\hat{C}_{\ell} & = & \sqrt{ \dfrac{\hat{C}_{\ell_{AB}}^{2} + 
\hat{C}_{\ell_{AA}}\hat{C}_{\ell_{BB}}}{2} }
\end{eqnarray}
and $\hat{C}_{\ell} \equiv  C_{\ell_{\rm ILC}} + N_{\ell_{\rm ILC}} = C_{\ell} + N_{\ell}$ is the sum of CMB, kSZ, residual foregrounds and noise in the two ILC maps $A$ and $B$. 

It is to be noted that although the cross-ILC combination returns a lower level of CIB+tSZ residuals compared to MV-ILC, modifying the ILC weights to suppress the foregrounds introduces a noise penalty.
We demonstrate this in Fig.~\ref{fig_appendix_bandpower_errors} in Appendix \ref{appendix_bandpower_errors}.
As a result the total kSZ SNR in the cross-ILC combination will be lower than the MV ILC. 
Despite this noise penalty, the low CIB+tSZ residuals in the cross-ILC combination will allow us to measure the kSZ signal robustly compared to the MV as discussed in Appendix~{\ref{appendix_bias_cibtsz}}.

\subsubsection{Fisher forecasts}
\label{sec_fisher_forecasts_for_ksz}

In Table~\ref{tab_total_ksz_snr} we present the Fisher forecasts for the total kSZ power spectrum $\dl = \aksz \times \kszleveldl\ \uksq$ for both the MV- and the cross-ILC combinations. 
The \snr{} is computed as $\sigma(\aksz)/\aksz$ and we marginalize over the 6 $\lcdm$ parameters and $\acibtsz$. 
Parameters governing the radio residuals, $\alpharad$ and $\alpharadsigma$, are fixed in this case but we also discuss below the impact on kSZ \snr{} when they are left free. 
We apply a \planck-like prior $\sigma(\taure)=0.007$ and $\sigma(\acibtsz)=0.1$ in this case. 
Like mentioned in \S\ref{sec_fisher_formalism}, we use $\ell \in [30, 5000]$ for T and P but modify $\ellmaxksz$ as shown in the table. 

As expected, the kSZ \snr{} increases when we include information from small-scales. 
This is because of the sample variance of the CMB which decreases exponentially when moving towards smaller scale due to diffusion damping. 
While the MV-ILC generally returns a better \snr{} compared to cross-ILC, having a precise knowledge of CIB and tSZ residuals is important for the MV-ILC (see Fig.~\ref{fig_cib_tsz_mvilc_residuals} and Fig.~\ref{fig_cib_plus_tsz_allilc_residuals} to compare the residual levels to kSZ). 
We demonstrate this in Appendix~{\ref{appendix_bias_cibtsz}} by calculating the biases in $\aksz$ estimation due to unmodeled CIB and tSZ signals. 
We can also note from the table, that the \snr{} saturates or even decreases for the MV-ILC when moving from $\ellmaxksz = 4000$ to $\ellmaxksz=4500$. 
This is because of the degeneracy between $\aksz$ vs $\acibtsz$ and the increase in the small-scale CIB+tSZ residuals. 
The case is different for cross-ILC and we note that the \snr{} constantly increases with $\ellmaxksz$. 
The choice of $\sigma(\acibtsz)=0.1$ prior does not affect the cross-ILC but has non-negligible impact on the results from MV-ILC. 

In the rest of the paper, we only focus on the constraints using the cross-ILC technique.
For \mbox{$\ellmaxksz = 4000\ (4500)$}, the kSZ power spectrum can be detected at $\sim19\sigma$ by \sptthreeg. 
Similarly, SO can measure the kSZ power spectrum at $13-15\sigma$.
The advantage of including HF bands for \sptthreeg{} and SO is also evident from the table. 
Including \sptfour{} will improve the \snr{} by $\sim \times2$ and this could be achieved during the later part of this decade. 
For SO, adding \ccatprime{} improves the \snr{} by $\sim \times1.5$. 
For \sfour, the kSZ SNR is $\sim 70\ (80)$ for \mbox{$\ellmaxksz = 4000\ (4500)$}. 
\\~\\
\noindent
{\bf \emph{Fitting for radio:}} When we include $\alpharad$ and $\alpharadsigma$, we see roughly 10\% reduction in kSZ \snr{} for $\ellmaxksz = 4000$. 
For, $\ellmaxksz = 4500$, the reduction in kSZ \snr{} is higher (30-50\%) compared to the baseline case when radio residual parameters are fixed. 
This is because of the degeneracy between $\aksz$ vs radio residuals when restricting the fitting to the same multipole ranges. 
If we include information from even smaller scales, the kSZ signal and radio residuals have significantly different shapes (see Fig.~\ref{fig_radio_residual_modelling}) which breaks the degeneracy between the parameters.
For example, when we set $\ellmax = 6000$, the reduction in kSZ \snr{} when fitting for radio residuals is lower only by 20\% compared to fixing the radio residual model parameters. 
The other way of mitigating this is by reducing the masking threshold or by nulling radio along with tSZ in one of the legs of the cross-ILC as described in \S~\ref{sec_radio_residuals} and demonstrated in Fig.~\ref{fig_radio_residual_modelling}.
The choice of the priors used for radio residuals (see Table~\ref{tab_parameters_priors}) does not have a significant impact on the kSZ \snr. 
For example, increasing the prior widths by $\times2$ changes the kSZ \snr{} only marginally.
\\~\\
\noindent
{\bf \emph{Impact due to marginalization of $\lcdm$ parameters:}} Assuming \planck{} priors on $\lcdm$ parameters has no impact on the kSZ \snr{} for \sfour. 
If all the $\lcdm$ parameters are fixed, the kSZ \snr{} increases kSZ by $20\%$. 
This is because the cosmological constraints are primarily driven by TE/EE for the future CMB surveys \citep{galli14} while kSZ constraints are TT-only. 
Alternatively, if we calculate the constraints with TT-only, then we find that applying \planck{} priors on $\lcdm$ parameters improves the kSZ \snr{} by 20\% while fixing them enhances the \snr{} by $\times2$.

\subsection{Constraints on the epoch of reionization}
\label{sec_eor_constraints}

\begin{figure*}
\centering
\ifdefined\ApJsubmit
\includegraphics[width=0.95\textwidth, keepaspectratio]{reionisation_constraints_ksz_tau_joint.pdf}
\else
\includegraphics[width=0.95\textwidth, keepaspectratio]{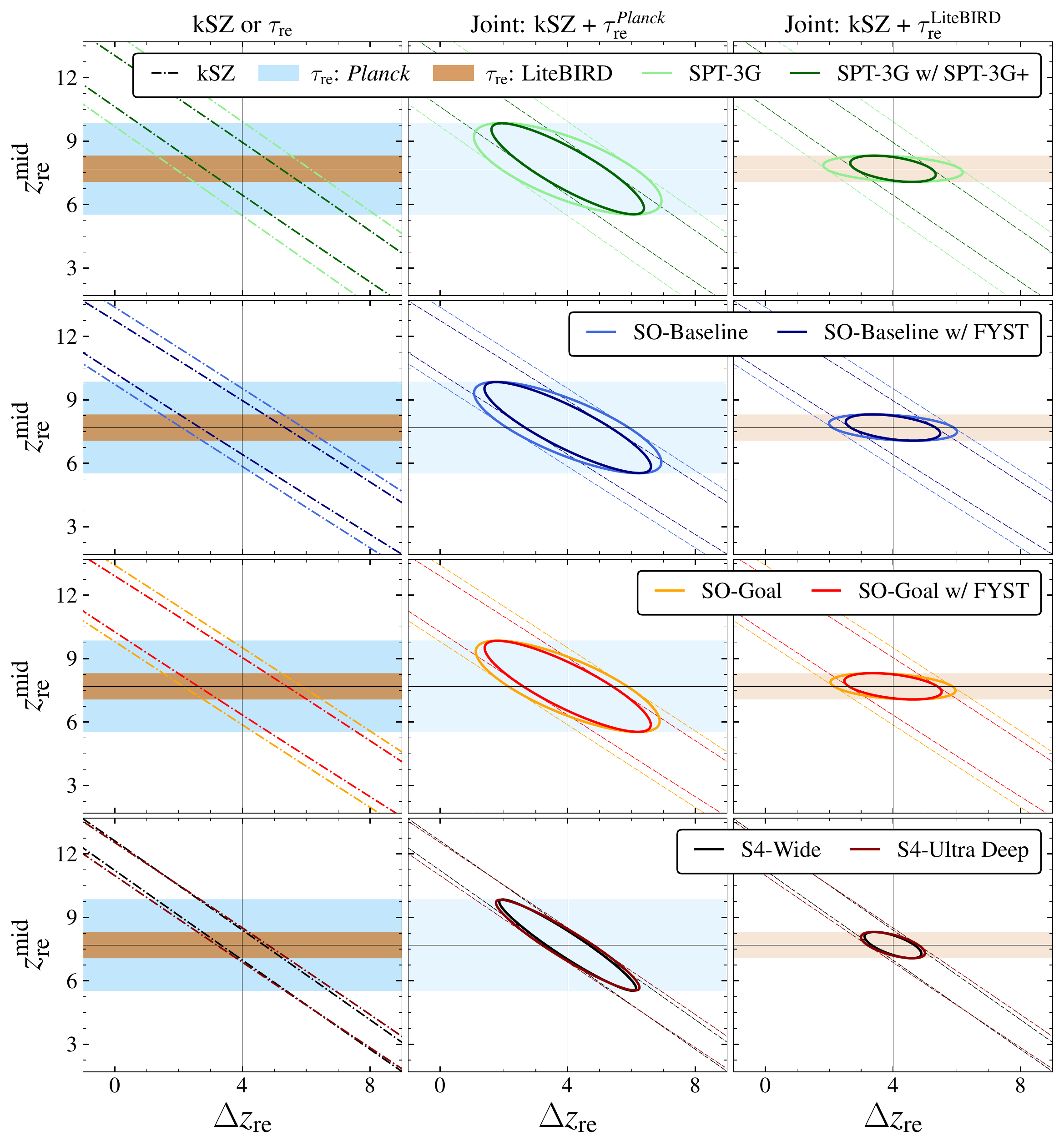}
\fi
\caption{Marginalized constraints on reionization parameters $\zmid$ and $\zdur$ from kSZ power spectrum (dash-dotted) and large scale E-mode measurements of $\sigma(\taure) = 0.007$ from \planck{} (blue shade) and $\sigma(\taure) = 0.002$ from \litebird{} (brown shade) in the left panel. 
The joint constraints from kSZ and \planck{} or \litebird{} are given in middle or right panels. 
kSZ power spectrum does not constrain $\zmid$ and hence it is dominated by the $\taure$ for all experiments. 
The future surveys can reduce the uncertainty on the duration of reionization $\sigma(\zdur) \sim 1.5$ (left panel) which is currently unconstrained by \planck.
The constraints on $\sigma(\zdur)$ improves by a $\times 2.5 - \times3$ when moving from \planck{} (middle panel) to \litebird{} (right panels) measurements of $\taure$.
}
\label{fig_reionisation_constraints}
\end{figure*}

In this section, we compute the constraints on reionization parameters from the kSZ power spectrum using the cross-ILC technique. 
Like mentioned in \S~\ref{sec_fisher_formalism}, we now replace the total kSZ amplitude $\aksz$ parameter with four other parameters namely: $\zmid, \zdur, \akszhomo, \alphakszhomo$ (see Table~\ref{tab_parameters_priors}). 
We use the standard priors listed in Table~\ref{tab_parameters_priors} for the foreground parameters: $\acibtsz, \alpharad, \alpharadsigma$.

To this end, we use Abundance Matching Box for the Epoch of Reionization (\amber) simulations \citep{trac22, chen23} to predict the patchy kSZ signal using two parameters: the mid-point $\zmid$ and duration $\zdur$ of reionization. 
Here \mbox{$\zdur \equiv \Delta z_{\rm dur, 90} = z_{\rm start} - z_{\rm end}$} where $z_{\rm start}$ and $z_{\rm end}$ correspond to the redshift where the Universe was 5\% and 95\% ionized, respectively.
The fiducial value for $\zmid = 7.69$, corresponding to $\taure = 0.0544$ \citep{planck20_2018cosmo}, and we set $\zdur = 4$ \citep{chen23}. 
Modifying the fiducial values of the two parameters by $\pm 1$, changes $\dl^{\rm kSZ}$ by $\sim20\%$ between $\ell \in [3000, 5000]$. \amber{} simulations also allows us to tweak the history of reionization to be either symmetric or asymmetric around the midpoint $\zmid$. This is parameterized using \mbox{$A_{z} = \dfrac{z_{\rm start} - z_{\rm mid}}{z_{\rm mid} - z_{\rm end}}$}. As can be seen, $A_{z} = 1$ indicates a symmetric reionization. \citet{chen23} showed that the variations in $A_{z}$ leads to small changes in the patchy kSZ power spectrum and measurement errors of $\lesssim 0.2 \uk^{2}$ are required across a wide range of multipoles to detect deviations in $A_{z}$ \citep[c.f. see right panel of Fig. 10 and top panel of Fig. 15 of ][]{chen23}. 
Subsequently, we choose to fix $A_{z} = 3$ and only vary $\zmid$ and $\zdur$.

For the late-time homogeneous kSZ signal, we follow \citep{alvarez20} and model it as \mbox{$\dl^{\rm h-kSZ} = \akszhomo \left(\dfrac{\ell}{\ell_{\rm \ast}}\right)^{\alphakszhomo} \uksq$} with $\akszhomo = 1.5$, $\alphakszhomo = 0$, and $\ell_{\rm \ast} = 3000$. 
We note here that this simple power-law formalism does not capture all the plausible kSZ late-time models. 
In particular, this parameterization ignores the correlation between the density field and the reionization history \citep{liu16} and hence the reionization constraints quoted below are slightly on the optimistic side.

Since it is hard to differentiate between the homogeneous and patchy kSZ signals from the total kSZ power spectrum, we set priors on them: $\sigma(\akszhomo) = 0.1$ and $\sigma(\alphakszhomo) = 0.1$ \citep{SO18}. 
We also discuss the impact on reionization constraints when the two above priors are softened.
There are a couple of other techniques of separating the kSZ contributions from the homogeneous and patchy components. 
First is the ``dkSZ-ing'' \citep{foreman22} approach of predicting and subtracting the homogeneous kSZ using large-scale structure (LSS) surveys. 
This is possible, since the homogeneous kSZ is sourced by haloes in the low redshift Universe which will he highly correlated with the LSS tracers. 
While the authors primarily focused on improving the constraints on cosmological parameters using dKSZ-ing, this technique should also be useful to extract the reionization kSZ signal, as alluded by authors in the paper. 
The second approach is to use kSZ 4-pt information \citep{smith17, alvarez20}. 
These are, however, outside the scope of this work and we leave these methods to be explored in a future work.

In Fig.~\ref{fig_reionisation_constraints}, we present the constraints on $\zmid$ and $\zdur$ after marginalizing over 6 $\lcdm$, 3 foreground, and 2 homogeneous kSZ parameters.
We show the constraints separately from kSZ (dash-dotted) and $\taure$ measurements (\planck{} as the blue shade and \litebird{} as brown shade) in the left panel. 
The joint constraints after combining kSZ with $\taure$ from \planck{} (\litebird) are shown in the middle (right) panels.
Since the kSZ power spectrum does not constrain $\taure$ or alternatively $\zmid$, they are dominated by $\taure$ for all experiments. 
We can note from the middle panel that \sptthreeg{} and SO can achieve $\sigma(\zdur) = 2$. 
Including HF information from \sptfour{} and \ccatprime{} (slightly darker curves), can reduce the uncertainty to $\sigma(\zdur) = 1.6$ while CMB-S4 can reduce it further to $\sigma(\zdur) = 1.4$. The constraints on $\sigma(\zdur)$ improves by a $\times 2.5 - \times3$ when moving from \planck{} (middle panel) to \litebird{} (right panels) measurements of $\taure$. 
We assume the $\taure$ measurements exclusively come from \planck{} and \litebird. This is a conservative choice since the errors on $\taure$ can reduce slightly when combining CMB and lensing power spectra.

Since the degeneracy between the two kSZ signals is important we also report the degradation in EoR constraints when adopting less constraining priors on the late-time kSZ signal. Widening the width of the priors $\sigma(\akszhomo)$ and $\sigma(\alphakszhomo)$ by $\times 2$ ($\times 5$) increases $\sigma(\zdur)$ by 10\% (30\%) for \sfour.

\subsection{Temperature-based CMB lensing}
\label{sec_cmb_lensing}

\begin{figure*}
\centering
\ifdefined\ApJsubmit
\includegraphics[width=0.84\textwidth, keepaspectratio]{lensing_xcorr_biases_s4wide_zbin3.pdf}
\else
\includegraphics[width=0.84\textwidth, keepaspectratio]{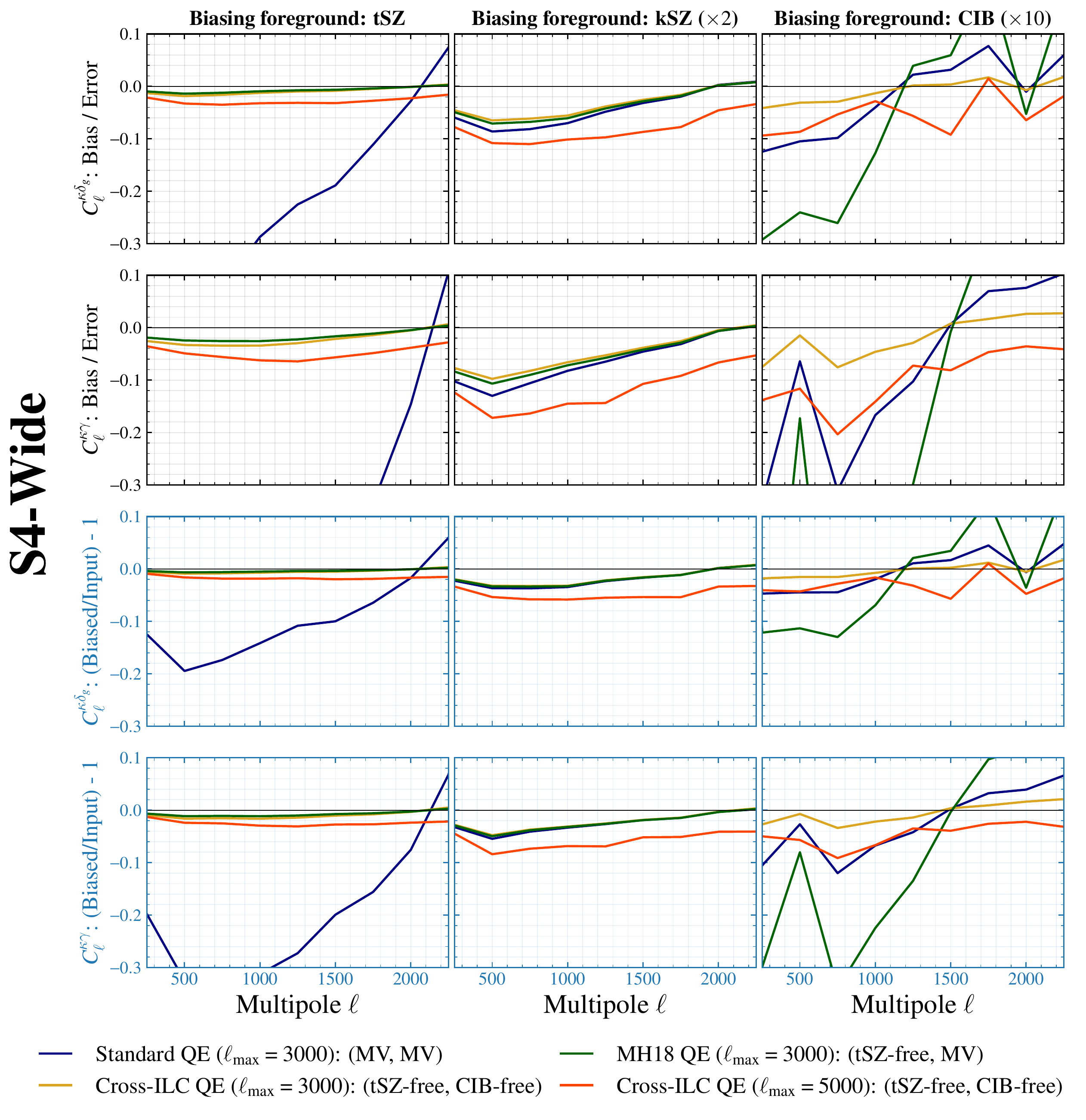}
\fi
\caption{
The observed foreground-induced lensing biases due to tSZ, kSZ, and CIB when cross-correlating CMB lensing convergence $\kappa$ map from \sfour{} with galaxy overdensity $\delta_{g}$ and shear $\gamma$ from LSST-Y1. The fiducial maximum CMB multipole is set to $\ellmax^{T} = 3000$ except for red the curves where we set $\ellmax^{T} = 5000$.
The standard QE \citep{hu02} (blue) is significantly affected by tSZ (left panel) while the other estimators are immune to tSZ. 
The bias form kSZ (middle panel) is the same for all estimators with $\ellmax^{T}=3000$ and small at $\lesssim 0.1\sigma$ ($\sim 1-3\%$) level. 
The bias from CIB follows this order: MH18 QE (green, \citetalias{madhavacheril18}) > standard QE (blue, \citealt{hu02}) > cross-ILC QE (yellow), but note that the CIB biases are boosted by $\times 10$ for clarity.
For all foregrounds, the bias from the cross-ILC is the lowest and hence we also quantify the bias for $\ellmax = 5000$ as well (red curve). 
As it is evident from the red curve, the foreground-induced biases are at a negligible level for cross-ILC QE even for $\ellmax^{T} = 5000$. 
We limit this figure to a single redshift bin $z \in [0.56, 82]$ but present the biases for multiple redshift bins in Fig.~\ref{fig_appendix_lensing_residuals}.
}
\label{fig_lensing_xcorr_biases_s4wide}
\end{figure*}

\citetalias{madhavacheril18} demonstrated that it is possible to produce temperature-based lensing maps that are immune to biases from tSZ with minimal noise penalty by utilizing two different maps: one high-resolution low-noise map (either the lowest noise frequency channel or the minimum-variance map) and a \tszfree{} map. One concern with such an approach is the boosted amplitude of CIB in the \tszfree{} map (see red curve in Fig.~\ref{fig_cib_plus_tsz_allilc_residuals}), which could be picked up strongly if the other input temperature map also contains CIB residuals.

In this study, we build on the general concept of \citetalias{madhavacheril18} and replace the low-noise/minimum variance temperature map with a \cibfree{} temperature map. 
This allows us to reduce biases from both tSZ and CIB simultaneously, allowing us to produce a clean CMB lensing map with minimal contamination from extragalactic foregrounds.

We follow the approach used in the original \agora{} simulations work \citep{omori22} and estimate the residual bias from various secondary components by performing lensing reconstruction on single component maps (i.e., individual temperature maps of tSZ/kSZ/CIB) processed in the same manner as the full lensing reconstruction (assuming the presence of all the foregrounds). 
Rather than showing the auto-spectrum of the reconstructed single-component lensing maps, we cross-correlate the reconstructed lensing maps with LSS tracers (galaxy overdensity and galaxy shear) since foreground biases get picked up more strongly in such cross-correlation measurements \citep{fabbian19}. 
Specifically, we use Rubin Observatory Legacy Survey of Space and Time (LSST-like, \citealt{lsst09}) sample for this test.


The results are shown in Fig.~\ref{fig_lensing_xcorr_biases_s4wide} where we compare biases from standard QE (blue), \citetalias{madhavacheril18} QE (green), and the cross-ILC QE (yellow and red). 
We show biases relative to the error $\dfrac{( \hat{C}_{\ell}^{AB} - C_{\ell}^{AB})}{\Delta C_{\ell}^{AB}}$ and also the differential bias $\dfrac{\hat{C}_{\ell}^{AB}}{C_{\ell}^{AB}}-1$. 
The errors are calculated analytically using Eq.~\ref{eq_knox_errors} where $A$ corresponds to CMB lensing $\kappa$ map from \sfour{} and $B$ is either the galaxy overdensity $\delta_{g}$ or shear $\gamma$ from LSST year 1 (LSST-Y1) sample.
We use CMB $\ellmax^{T} = 3000$ in all cases except for the red curve for which we include information up to $\ellmax^{T} = 5000$.  
We perform this cross-correlation measurement in multiple redshift bins for different experiments although in Fig.~\ref{fig_lensing_xcorr_biases_s4wide} we only show the results for the redshift bin $z \in [0.56, 0.82]$ for \sfour{} for brevity. 
The results for other experiments and all the redshift bins are presented in Fig.~\ref{fig_appendix_lensing_residuals} in Appendix~\ref{appendix_lensing_biases}. 

From Fig.~\ref{fig_lensing_xcorr_biases_s4wide}, we note that the standard QE (blue) shows significant biases due to tSZ (left panel) as expected. 
The other curves are immune to tSZ bias. 
Note that the tSZ biases are non-zero due to the finite band passes of the experiments. See Appendix~\ref{appendix_lensing_biases} for more details. 
The bias from kSZ (boosted by $\times2$ for clarity) is also small $\lesssim 0.1\sigma$ ($\sim 1-3\%$) for the lensing estimators. 
The bias from CIB is higher for \citetalias{madhavacheril18} QE (green) due to the enhanced level of CIB in the \tszfree{} map. 
However, note that the CIB bias is boosted by $\times 10$. 
For cross-ILC QE, the bias from CIB is also negligible. 

\section{Conclusion}
\label{sec_conclusion}

We presented a cross-ILC approach to robustly extract the kSZ and CMB lensing signals from the current (\sptthreeg) and future (SO and \cmbsfour) CMB surveys. 
The approach uses the cross-spectrum between \cibfree{} and \tszfree{} cILC maps for kSZ measurement. 
For CMB lensing, we pass the \cibfree{} and \tszfree{} cILC maps in the two legs of the QE. 

We showed that the residual bias from CIB and tSZ is minimal for this approach and $\times3$ to $\times5$ lower compared to the expected kSZ signal level of $D_{\ell} = \kszleveldl\ \mu {\rm K}^{2}$. 
The CIB and tSZ residuals in this cross-ILC combination are also significantly lower than the measurements using auto-spectrum of either the MV or the cILC map. 
We also quantified the residual radio signals and demonstrated that they can be modeled or mitigated easily. 
We have ignored galactic foregrounds in this work as they are negligible on small scales compared to the extragalactic foreground signals. 
Similarly we have also ignored CO in this work as they are also much smaller compared to kSZ and other foreground signals \citep{maniyar23}.
For CMB lensing, using cross-correlation with galaxy overdensity and shear fields from LSST-Y1 sample, we showed that the modified cross-ILC QE has negligible level of CIB and tSZ-induced biases. 

Using Fisher formalism, we forecasted the expected kSZ \snr{} for CMB surveys.
For the cross-ILC technique, we showed that the total kSZ power spectrum can be detected at $\sim 19\sigma$ by \sptthreeg{} and the inclusion of data from its successor \sptfour, improves the \snr{} by $\times2$. 
The expected kSZ \snr{} for SO is $\sim13-15\sigma$ with roughly $\times1.5$ expected after the inclusion of information from \ccatprime. 
For CMB-S4, the expected \snr{} of the total kSZ power spectrum is extremely high $70-80\sigma$. 

We also estimated the biases due to unmodeled CIB/tSZ residuals and found that the kSZ measurements from the MV-ILC can be significantly biased. 
The biases on the cosmological parameters are negligible, though, since the constraints are dominated by EE/TE. 
For the cross-ILC measurement, biases on both kSZ and cosmological parameters are negligible. 

We also forecasted the constraints on the epoch of reionization, $\zmid$ and $\zdur$ from kSZ power spectrum using the cross-ILC approach. 
While the error on $\zmid$ is dominated by the prior on the optical depth, the current and upcoming surveys can achieve $\sigma(\zdur) = 1.4-2$. This improves by $\times2.5-\times3$ when we replace the prior on optical depth from \planck-like to \litebird-like. 
Including the kSZ 4-pt information can improve these constraints further \citep{smith17}.
\\~\\
\noindent
{\it Applications:} 
The technique has several potential applications. 
For example, it can also be used to mitigate the effects of dust and synchrotron signals, both in temperature and polarization, using dust-/synchrotron-free maps in the two legs of the cross-ILC estimator facilitating robust measurements of large-scale CMB for inflationary B-modes. 
It can also be used for other higher order statistics by plugging in different foreground-removed ILC maps in each of the legs. For example, projected field kSZ \citep{hill16, kusiak21} as explored recently by \citet{kusiak21}, kSZ 4-pt \citep{smith17, ferraro18, alvarez20}, tSZ bispectrum analysis \citep{crawford14}. 
We have not shown these explicitly here and leave them to be explored in a future work. 
\\~\\
\noindent
{\it Data/code availability:} 
All the data products produced in this work and the associated codes are publicly available and can be downloaded from this \href{https://github.com/sriniraghunathan/cross_ilc_methods_paper}{link$^{\text{\faGithub}}$}.

\section*{Acknowledgments}
We thank Gil Holder for several insightful discussions throughout the course of this work. 
We thank Tom Crawford for important discussions, constructive comments, and helpful suggestions that had significantly improved this manuscript. 
We also thank Alex Van Engelen, Colin Hill, Nicholas Huang, Abhishek Maniyar, Christian Reichardt, Nathan Whitehorn and Kimmy Wu for discussions and feedback on the manuscript. 
Finally, we thank the anonymous referee for helpful suggestions that helped in shaping this manuscript better.

This work was partially supported by the Center for AstroPhysical Surveys (CAPS) at the National Center for Supercomputing Applications (NCSA), University of Illinois Urbana-Champaign. 
This work made use of the following computing resources: Illinois Campus Cluster, a computing resource that is operated by the Illinois Campus Cluster Program (ICCP) in conjunction with the National Center for Supercomputing Applications (NCSA) and which is supported by funds from the University of Illinois at Urbana-Champaign; the computational and storage services associated with the Hoffman2 Shared Cluster provided by UCLA Institute for Digital Research and Education's Research Technology Group; and the computing resources provided on Crossover, a high-performance computing cluster operated by the Laboratory Computing Resource Center at Argonne National Laboratory.
\appendix
\restartappendixnumbering

\section{Bandpower error comparisons}
\label{appendix_bandpower_errors}

\begin{figure*}
\centering
\ifdefined\ApJsubmit
\includegraphics[width=
0.9\textwidth, keepaspectratio]{bandpower_errors.pdf}
\else
\includegraphics[width=0.9\textwidth, keepaspectratio]{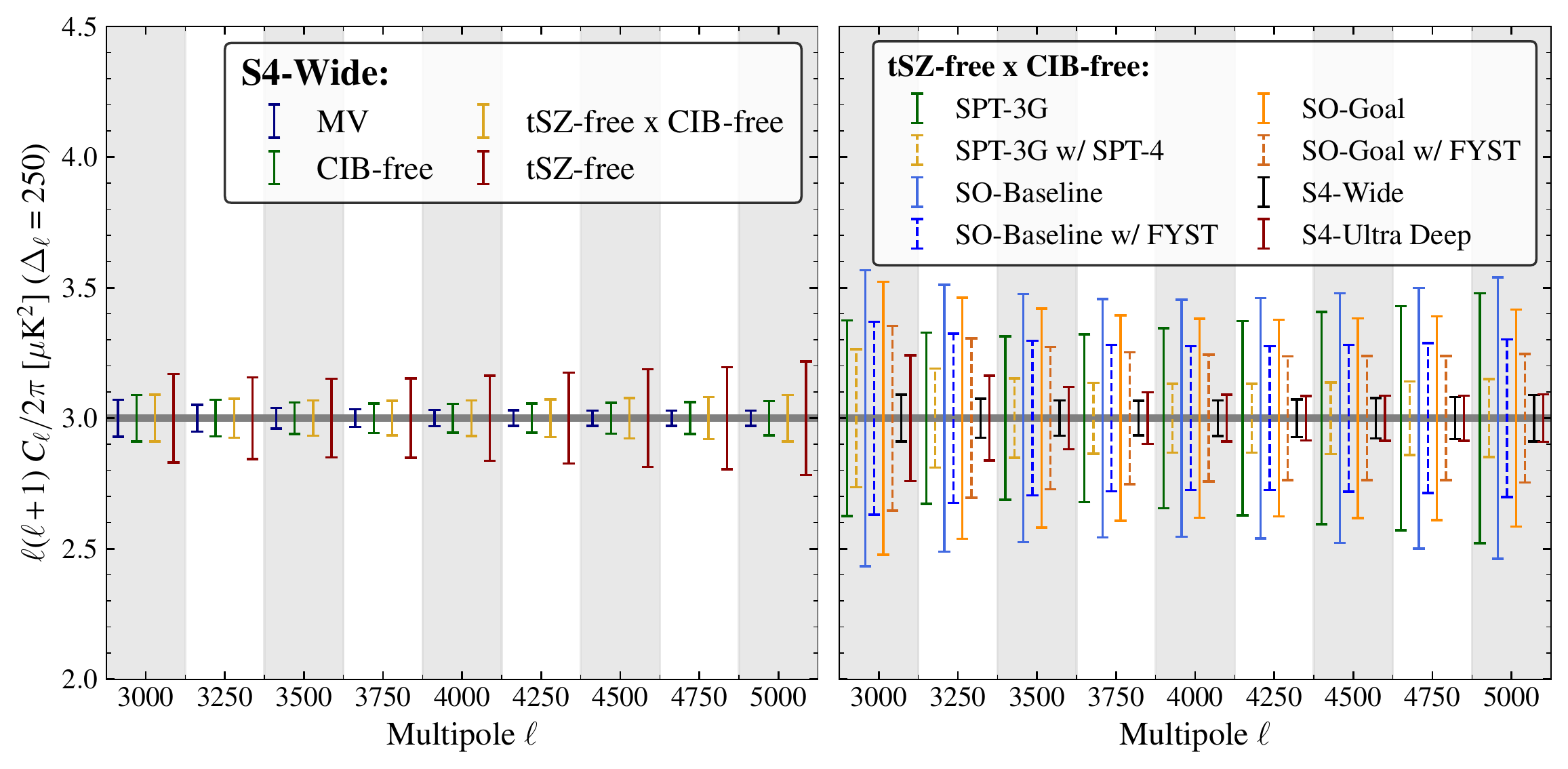}
\fi
\caption{Bandpower errors for \sfour{} survey from different ILC techniques is shown in the left panel. The data points include contribution from CMB, kSZ, residual foregrounds and noise. 
As expected, MV returns the least error while \tszfree{} maps has the largest error. The cross-ILC combination \mbox{\tszfree{} $\times$ \cibfree} is $\sim \times2$ worse (better) than MV (\tszfree). 
The right panel shows the bandpower errors expected from the cross-ILC technique for different experiments considered in this work. 
We assume $\Delta_{\ell} = 250$ for this figure. 
From the right panel, it is evident that the cumulative kSZ \snr{} for all experiments is high.
}
\label{fig_appendix_bandpower_errors}
\end{figure*}
We show the bandpower errors from different ILC techniques for \sfour{} in the left panel of Fig.~\ref{fig_appendix_bandpower_errors}. 
The cross-ILC technique is worse (better) than MV (\tszfree) by $\sim \times2$ across the entire multipole range shown in the figure. 
In the right panel, we show the errors for different experiments considered in this work from the cross-ILC  (\tszfree{} $\times$ \cibfree) technique. 
We assume $\Delta_{\ell} = 250$ to compute the bandpower errors using Eq.~\ref{eq_knox_errors}. 
We note that the cumulative kSZ \snr{} expected from all experiments is high with the cross-ILC technique. 
As shown previously in \S\ref{sec_total_cib_tsz_residuals} and Fig.~\ref{fig_cib_plus_tsz_residuals_crossilc_all_experiments}, the foreground residuals are lower for the cross-ILC approach making it optimal for extracting the kSZ signal from current and future experiments.

\section{Bias due to residual CIB and tSZ signals}
\label{appendix_bias_cibtsz}

\begin{figure*}[h]
\centering
\ifdefined\ApJsubmit
\includegraphics[width=
0.9\textwidth, keepaspectratio]{aksz_bias_due_to_cibtsz.pdf}
\else
\includegraphics[width=0.9\textwidth, keepaspectratio]{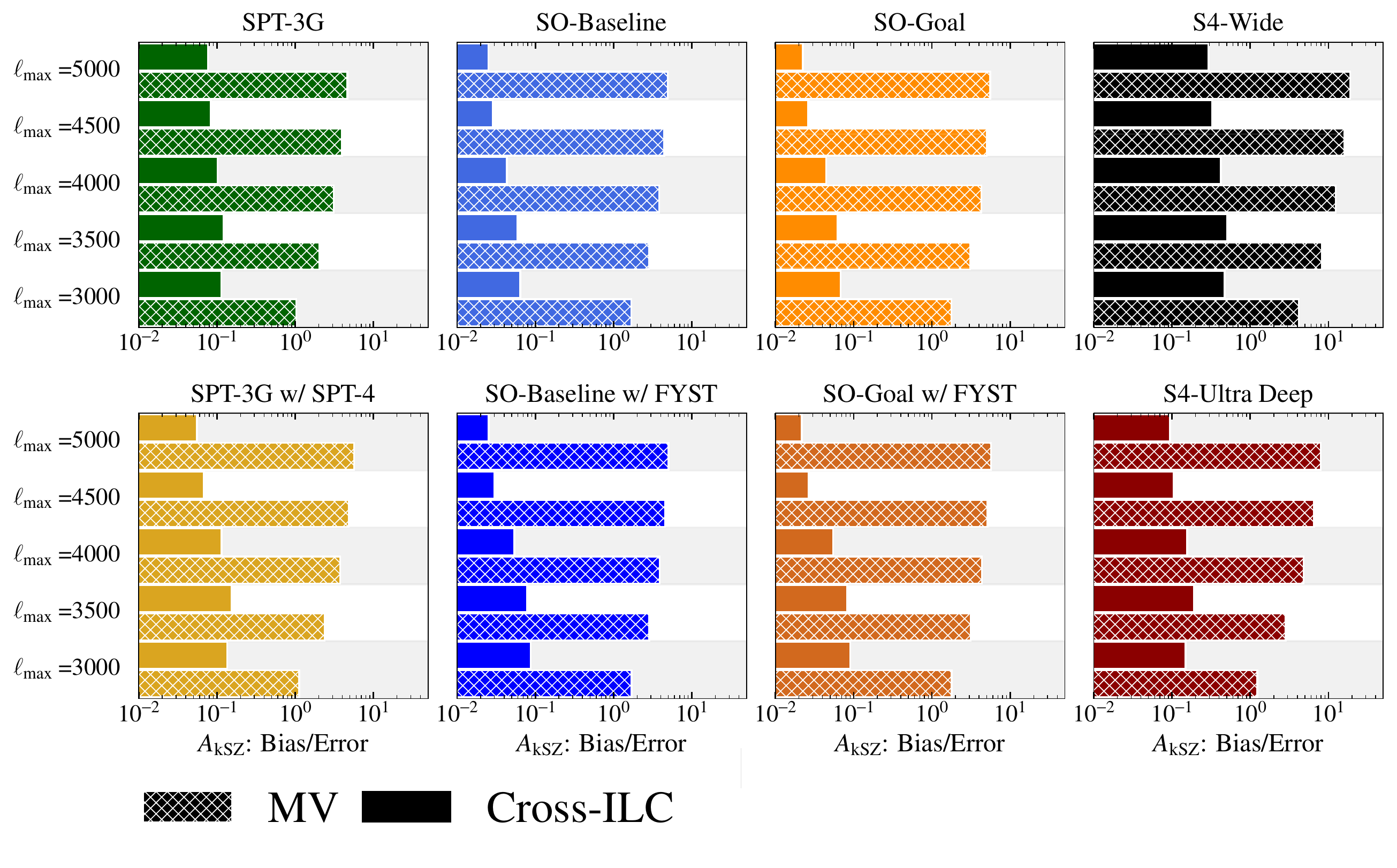}
\fi
\caption{Systematic biases in kSZ power spectrum estimation $\sigma_{\rm sys}(\aksz)$ due to 5\% unmodeled CIB and tSZ residuals in the CMB maps. The bias in the cross-ILC technique is low compared to MV, which is significantly biased. 
As expected, the bias in MV increases when extending $\ellmax$ in the analysis.
}
\label{fig_appendix_ksz_bias_cibtsz}
\end{figure*}

\begin{wrapfigure}{rb}{0.48\textwidth}
\vskip -30pt
\hspace{-0.7cm}
\ifdefined\ApJsubmit
\includegraphics[width=0.47\textwidth, keepaspectratio]{cosmo_params_bias_due_to_cibtsz_s4_wide.pdf}
\else
\includegraphics[trim=0.in 0.in 0in 0.0in,clip=true, width=0.47\textwidth, keepaspectratio]{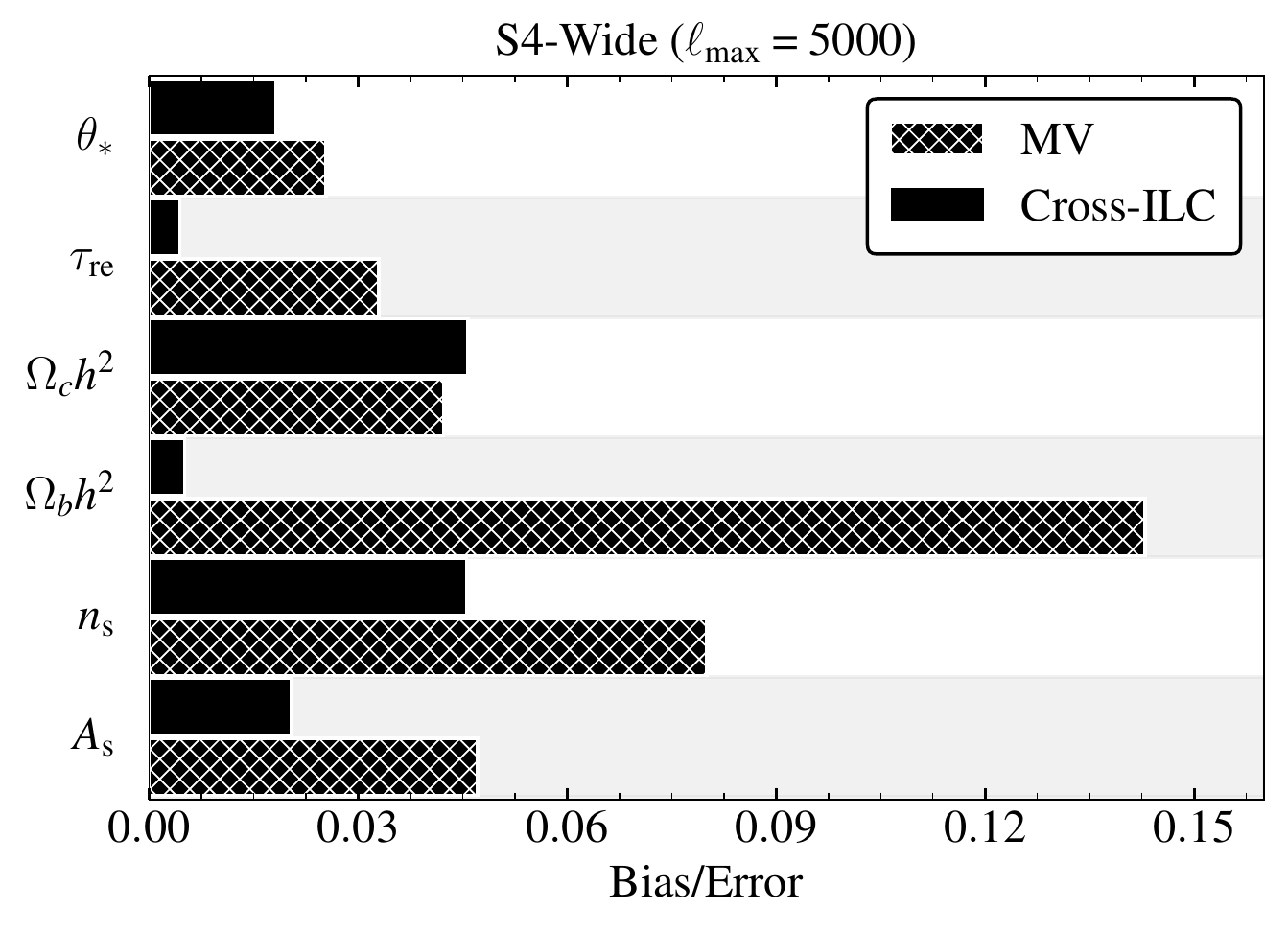}
\fi
\caption{Biases in cosmological parameters due to residual CIB and tSZ signals for \sfour{} survey. The biases are negligible for both MV and cross-ILC since the constraints are dominated by EE/TE rather than TT.}
\vskip -35pt
\label{fig_appendix_cosmo_bias_cibtsz}
\end{wrapfigure}
In this section we calculate the biases in $\aksz$ due to 5\% unmodeled residual CIB and tSZ signals in the CMB maps. 
We use a modified Fisher formalism as given by Eq.(\ref{eq_fisher_bias}) based on \citet{amara08} for estimating the biases. 
The bias $b$ on the parameter $i$ is calculated as
\begin{eqnarray}
b_{i} = F^{-1}_{ij} B_{j}\\
{\rm with\ }
B_{j} = \sum_{\ell} C_{\ell}^{\rm sys}\Delta C_{\ell}^{-2} \frac{\partial C_{\ell}^{XY}}{\partial{\theta_{j}}}.
\label{eq_fisher_bias}
\end{eqnarray}
Here $\Delta C_{\ell}$ is calculated with Eq.(\ref{eq_knox_errors}) but after including the systematic signal as well with $\hat{C}_{\ell} = C_{\ell} + N_{\ell} + C_{\ell}^{\rm sys}$.
The results are shown in Fig.~\ref{fig_appendix_ksz_bias_cibtsz} for both MV (hatched bars) and cross-ILC techniques (filled bars) as a function of $\ellmax$ used for kSZ extraction. 
We note from the figure that the MV ILC becomes significantly biased due to the residual CIB and tSZ signals while the biases in the cross-ILC technique are low in all cases. 
As expected, biases in MV ILC increases when the $\ellmax$ is extended since the level of CIB+tSZ residuals also increase with $\ellmax$ as can be seen from the blue curve of Fig.~\ref{fig_cib_plus_tsz_allilc_residuals}.  

We also calculate the biases in cosmological parameters due to the CIB/tSZ residuals and do not find significant biases. This lower level of biases is because the cosmological constraints are dominated by EE/TE while CIB/tSZ residuals are expected to be largely unpolarized and hence only injected for TT only. 
The biases in cosmological parameters expected due to residual foregrounds for \sfour{} are shown in Fig.~\ref{fig_appendix_cosmo_bias_cibtsz}.

\section{Validating \agora{} foreground model}
\label{appendix_validation_fg_model}
\subsection{SPT foreground model for ILC weights}
\label{appendix_validation_spt_fg_model}

We have used \agora{} simulations as our baseline foreground model. 
In this section we test the robustness of this assumption.

In the first test, we tweak the \agora{} CIB model by drawing random samples for CIB parameters from $1\sigma$ and $2\sigma$ regions from Fig.6 of \citet{omori22}. This leads to a negligible change in the final CIB and tSZ residuals for the cross-ILC technique, where the residuals are more than an order of magnitude smaller than the expected level of the kSZ signal.

In the second test, we fully modify the \agora{} foreground model. To this end, replace the $\rm N_{ch} \times \rm N_{ch}$ covariance matrix ${\bf C}_{\ell}$ in Eq.(\ref{eq_ilc_weights}) containing the auto- and cross-power spectra between \agora{} simulations in different bands by the foreground model from SPT measurements \citepalias{reichardt21}. 
Modifying the covariance matrix alters the ILC weights which will indeed modify the final residuals. 
We pass the \agora{} simulations through these modified weights and compare the resultant residuals with our baseline approach.

\begin{wrapfigure}{rb}{0.55\textwidth}
\vskip -10pt
\ifdefined\ApJsubmit
\includegraphics[trim=0.in 0.in 0in 0.0in,clip=true, width=0.54\textwidth, keepaspectratio]{ilc_cib_plus_tsz_residuals_agora_vs_spt.pdf}
\else
\includegraphics[trim=0.in 0.in 0in 0.0in,clip=true, width=0.54\textwidth, keepaspectratio]{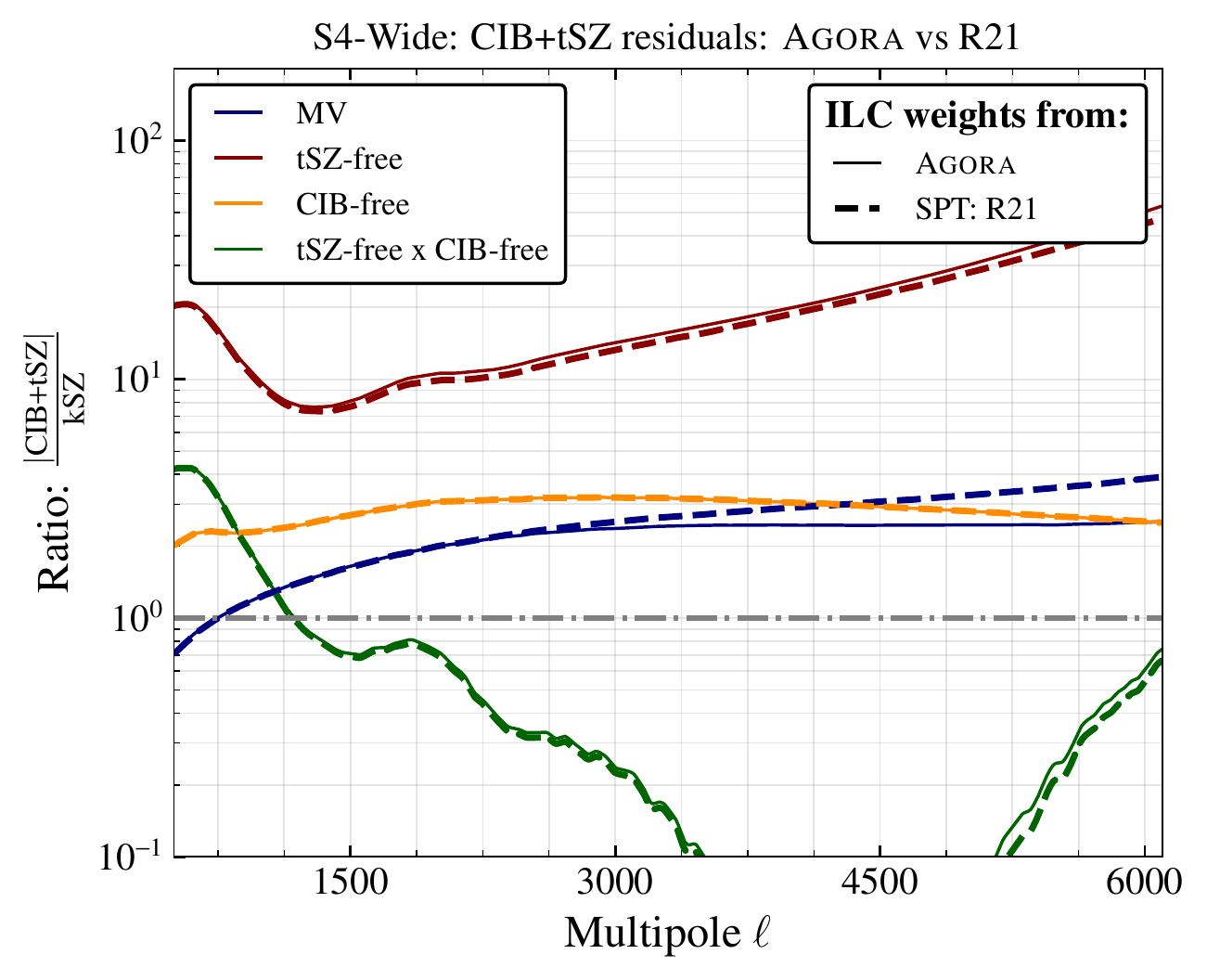}
\fi
\caption{Total CIB and tSZ residuals from \agora{} simulations for two different ILC weights. Solid lines correspond to the case when \agora{} foreground model is used in the covariance matrix to calculate the weights while the dashed lines use a foreground model based on SPT measurements. The colors correspond to different ILC combinations. 
The close match between the solid and dashed curves asserts that our baseline approach of using \agora{} foreground model is a valid assumption.}
\label{fig_fgmodel_agora_spt}
\end{wrapfigure}

Fig.~\ref{fig_fgmodel_agora_spt} presents this comparison for different ILC combinations: MV (blue), \tszfree{} (red), \cibfree{}  (orange) and the cross-ILC \tszfree{} $\times$ \cibfree{} (green). 
The solid lines correspond to ILC weights from \agora{} foreground model (baseline case) and the dashed lines are for SPT foreground model. 
All curves show the total CIB+tSZ foreground residuals. 

It is important to point out that there are some differences between SPT and \agora{} foreground model used in this work.
Firstly, the SPT foreground model computed using \citetalias{reichardt21} measurements does not have cluster masked and the masking threshold for dusty point sources in $S_{150} = 6.4\ {\rm mJy}$ while for \agora{} model, we have masked clusters with mass $\mvir \ge  \clustermask\ \msol$ and use a reduced masking threshold for dusty  sources $S_{150} = 3\ {\rm mJy}$.
For radio point sources, both the models assume the same $S_{150} = 3\ {\rm mJy}$ threshold. 
Next, the cross-correlation spectra between CIB and tSZ is simply modeled using a fixed cross-correlation coefficient of $\rho_{\rm CIB \times tSZ} = 0.078$ \citepalias{reichardt21} in the SPT model. 
Finally, the SPT CIB model is an extrapolation of the 150 GHz power spectrum assuming two SEDs (one for the Poisson component and the other for the clustering component of the CIB) and does not account for the CIB decorrelation between different frequency bands. 
As a result, we expect some differences between the solid and dashed curves. 

In Fig.~\ref{fig_fgmodel_agora_spt}, the \tszfree{} (red) curve is sensitive to the differences in the CIB and radio models between \agora{} and SPT while the \cibfree{} (orange) curve is sensitive to the differences in the tSZ and radio models.
The cross-ILC \tszfree{} $\times$ \cibfree{} (green) and the MV (blue) curves are affected by all the three (CIB, radio, and tSZ) foreground modeling differences.
Despite the above mentioned differences between the two foreground models, the agreement between the solid and dashed curves is excellent in all cases. 
Specifically, the difference is negligible for cross-ILC \tszfree{} $\times$ \cibfree{} (green). 
Based on this result, we claim that our baseline approach of using \agora{} simulations to estimate the ILC covariance matrix is reasonable. 
We also note that for future CMB surveys, the ILC covariance matrix can use the bandpowers measured by the respective experiment to derive the ILC weights rather than \agora.

\subsection{\websky{} foreground simulations}
\label{appendix_validation_websky}
Now we test the cross-ILC approach by replacing \agora{} using the multi-component Websky\footnote{Websky simulations are downloaded from \url{https://mocks.cita.utoronto.ca/index.php/WebSky_Extragalactic_CMB_Mocks}} simulations \citep{stein20}. 
Similar to \S~\ref{appendix_validation_spt_fg_model}, we limit this test to the \sfour{} survey. 
We approximate $\nu \in [90, 150, 220, 285]$ GHz with the publicly available [93, 145, 217, 278] GHz Websky simulations and do not attempt to interpolate the simulations to match the bands used in this work. 
We also note that the Websky simulations assume a delta function for the bandpasses centred on each frequency band \citep{stein20}. 
Even though the CIB power in Websky simulations are shown to match \planck{} 545 GHz data well \citep{stein20}, the CIB power at $\ell = 3000$ in the 150 GHz band is $\sim \times2$ higher \citep{raghunathan22} than the SPT results reported by \citep{george15, reichardt21}. Subsequently, we multiply the Websky CIB maps by 0.75. 
The applied correction is the same for all bands. 
We note this simple scaling alone does not make Websky CIB to match SPT results in other bands but it it makes the agreement better. 
This simple $\ell$-independent scaling also does not work for the large-scale clustered part of the CIB signal. 
No scaling is applied for Websky tSZ maps. 

With the Wesbky CIB/tSZ maps and the power spectra in hand, we run the blind search algorithm described in \S~\ref{sec_cross_ilc} to derive the optimal weights for the cross-ILC estimator. 
We note that the SEDs obtained from the blind search does not match the SEDs obtained for \agora{} simulations. 
This is because of the differences between Websky and \agora{} simulations, although we find \agora{} to match data better as demonstrated in Fig.~\ref{fig_fgmodel_agora_spt}. 
Nevertheless, adopting the SEDs from the blind search does reduce the total CIB+tSZ residuals. 
For example, without the Websky CIB scaling mentioned above, the total CIB+tSZ residuals are
$\times1.5$ lower than kSZ between $\ell \in [3000, 5000]$.
However, after introducing the CIB scaling, we find the CIB+tSZ residuals to go down by $\times2$ at $\ell = 3000$ and by $\times4$ $\in [4000, 5000]$ compared to the expected level of the kSZ signal $\dl = \kszleveldl \uksq$. 

\section{Lensing biases}
\label{appendix_lensing_biases}

\begin{figure*}
\centering
\ifdefined\ApJsubmit
\includegraphics[width=0.95\textwidth, keepaspectratio]{lensing_xcorr_biases_allexps.pdf}
\else
\includegraphics[width=0.95\textwidth, keepaspectratio]{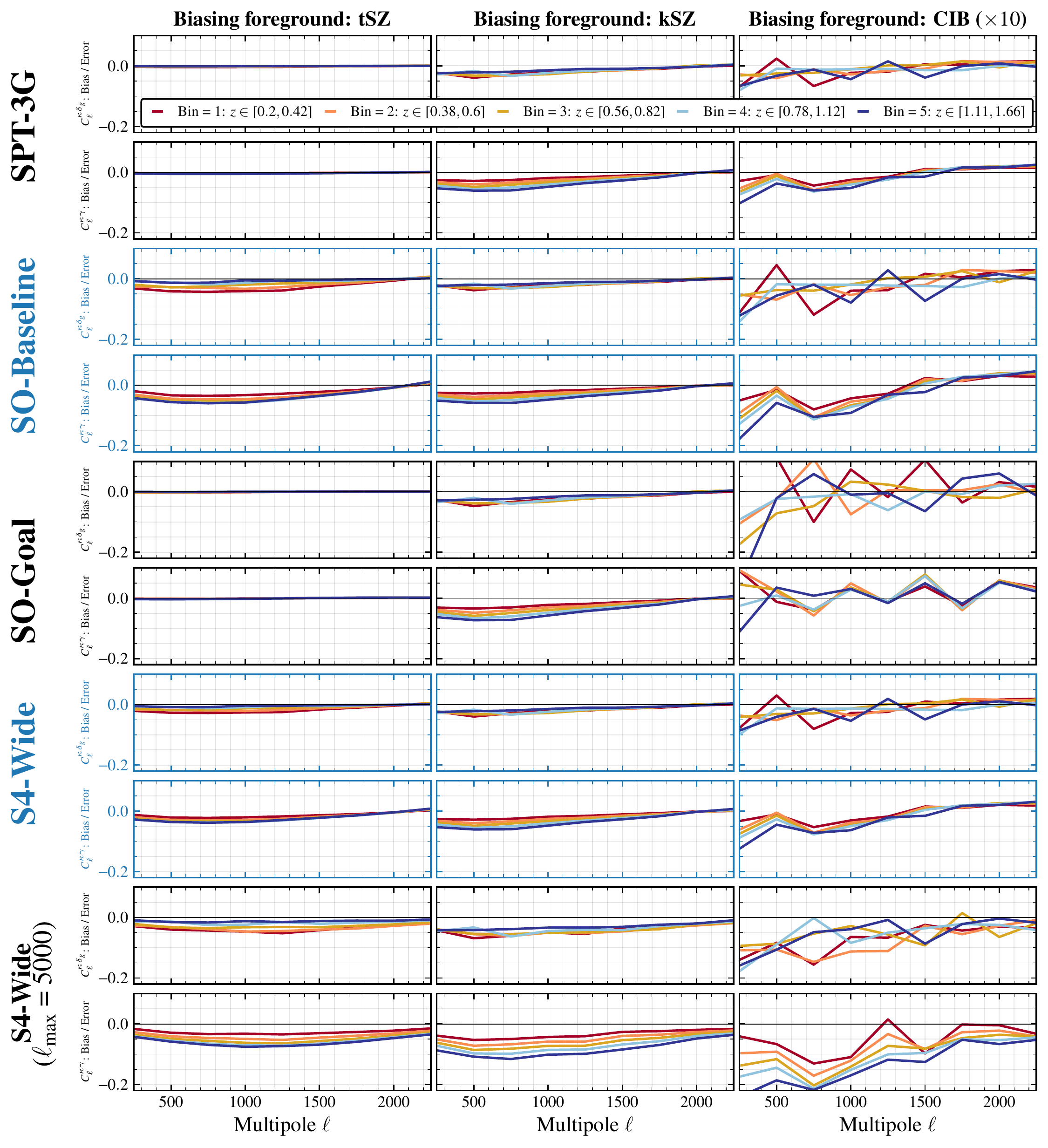}
\fi
\caption{Biases due to tSZ (left panel), kSZ (middle panel), and CIB (right panel) in CMB lensing cross-correlations similar to Fig.~\ref{fig_lensing_xcorr_biases_s4wide} but extended to other experiments and more redshift bins. See text for more details.}
\label{fig_appendix_lensing_residuals}
\end{figure*}

In Fig.~\ref{fig_appendix_lensing_residuals}, we show the biases in CMB lensing cross-correlations for the cross-ILC QE. 
This is similar to Fig.~\ref{fig_lensing_xcorr_biases_s4wide} but here, along with \sfour, we
also include results for \sptthreeg{}, \sofid, and \sogoal. 
In this figure, we only show the bias over the error.
We also present the results for five redshift bins: Bin 1 = $z \in [0.2, 0.42]$, Bin 2 = $z \in [0.38, 0.6]$; Bin 3 = $z \in [0.56, 82]$; Bin 4 = $z \in [0.78, 1.12]$; and Bin 5 = $z \in [1.11, 1.66]$. 
Note that the redshift bins are not independent and have some overlap due to photo-$z$ errors \citep{lsstdesc18} although that is irrelevant in this context since we are mainly focused on the foreground-induced biases in CMB lensing cross correlations. 
The biases are all calculated using \agora{} simulations \citep{omori22}.

The bias due tSZ (left panel) is small $\sim 0.05\sigma$ but non-zero in some cases. 
This may be a little counter intuitive since the cross-ILC lensing QE uses the \tszfree{} map in one of the legs and we should expect the bias from tSZ to be zero. 
However, note that the finite band passes of the experiments can lead to small levels of residual tSZ and that can correlate with the non-zero tSZ signal in the \cibfree{} map. 
Nevertheless, the level of the tSZ bias is negligible. 
The bias from kSZ is also at a similar level for all experiments. 
The bias due to CIB is at sub-percent level and hence negligible. 
Note that we have boosted the CIB biases by $\times10$ for clarity.

Motivated by these small level of biases, in the bottom panel we present the results for \sfour{} but after increasing the CMB $\ellmax^{T} = 5000$ during lensing reconstruction. 
As discussed earlier in \S\ref{sec_cmb_lensing}, including information from smaller scales increases the lensing SNR significantly. 
For example, the lensing SNR with the maximum CMB multipole $\ell_{\rm max}^{T} = 5000$ is roughly $\times 1.7$ better than using $\ell_{\rm max}^{T} = 3000$ for temperature-based CMB lensing.
This can also lead to enhanced level of biases which we test here. 
However, as can be seen from the bottom panel, the biases are small 
$\lesssim 0.1\sigma$
even in this case. 
While we do not show it explicitly, we also checked the biases due to radio sources for the fiducial masking threshold of $S_{150}$ = 6 mJy and find them also to be at sub-percent level.
Thus, we conclude that the cross-ILC lensing QE is robust against the biases induced by CIB, kSZ, radio and tSZ signals.

\bibliography{lensing_sz}
\end{document}